\documentclass[11pt,oneside]{article}
\usepackage[margin=1.0in]{geometry}
\usepackage{amsmath}
\usepackage{amsfonts}
\usepackage{amsthm}
\usepackage{enumerate}
\usepackage[T1]{fontenc}
\usepackage{mathrsfs}
\usepackage{mathdots}
\usepackage{graphicx}
\usepackage{mathtools}
\graphicspath{ {./images/} }
\usepackage{xcolor}
\usepackage{hyperref}
\hypersetup{colorlinks=true,linkcolor=red,citecolor=blue,filecolor=magenta,urlcolor=cyan}
\usepackage{subfiles}
\usepackage{pgfplots}
\usepackage{tikz}
\usepackage{algpseudocode}
\usepackage{algorithm}

\usepackage[utf8]{inputenc}

\usepackage[
backend=biber,
style=alphabetic,
sorting=nyt
]{biblatex}

\newtheorem{theorem}{Theorem}[section]

\newtheorem{lemma}[theorem]{Lemma}

\theoremstyle{definition}
\newtheorem{remark}[theorem]{Remark}
\newtheorem{definition}[theorem]{Definition}

\newcommand{\Z}{\mathbb{Z}}

\newcommand\numberthis{\addtocounter{equation}{1}\tag{\theequation}}
\numberwithin{equation}{section}

\newcommand{\Otil}{\tilde{O}}
\newcommand{\Ohat}{\hat{O}}
\DeclareMathOperator{\nnz}{nnz}

\newtheorem{innercustomgeneric}{\customgenericname}
\providecommand{\customgenericname}{}
\newcommand{\newcustomtheorem}[2]{%
  \newenvironment{#1}[1]
  {%
   \renewcommand\customgenericname{#2}%
   \renewcommand\theinnercustomgeneric{##1}%
   \innercustomgeneric
  }
  {\endinnercustomgeneric}
}

\newcustomtheorem{customthm}{Theorem}
\newcustomtheorem{customlemma}{Lemma}
\newcustomtheorem{customdef}{Definition}
\newcustomtheorem{customcor}{Corollary}

\title{\vspace{0.7cm} Faster Sparse Matrix Inversion and Rank Computation in Finite~Fields}

\author{
S\'\i lvia Casacuberta\thanks{This work was partly done while S. Casacuberta was at ETH Zurich, funded by the startup grant of Prof.~Kyng.}  \\ \texttt{scasacubertapuig@college.harvard.edu} \\ Harvard University
\and
Rasmus Kyng \\ \texttt{kyng@inf.ethz.ch} \\ ETH Zurich
}

\date{}

\pgfplotsset{compat=1.17}

\begin{document}

\maketitle

\thispagestyle{empty}

\begin{abstract}
We improve the current best running time value to invert sparse matrices over finite fields, lowering it to an expected $O\big(n^{2.2131}\big)$ time for the current values of fast rectangular matrix multiplication. 
We achieve the same running time for the computation of the rank and nullspace of a sparse matrix over a finite field. This improvement relies on two key techniques. First, we adopt the decomposition of an arbitrary matrix into block Krylov and Hankel matrices from Eberly et al.\ (ISSAC 2007). 
Second, we show how to recover the explicit inverse of a block Hankel matrix using low displacement rank techniques for structured matrices and fast rectangular matrix multiplication algorithms. 
We generalize our inversion method to block structured matrices with other displacement operators and strengthen the best known upper bounds for  explicit inversion of block Toeplitz-like and block Hankel-like matrices, as well as for explicit inversion of block Vandermonde-like matrices with structured blocks. 
As a further application, we improve the complexity of several
algorithms in topological data analysis and in finite group theory.
\end{abstract}

\newpage

{
  \hypersetup{linkcolor=black}
  \tableofcontents
}

\thispagestyle{empty}

\newpage

%

\pagestyle{plain}
\setcounter{page}{1}

\section{Introduction}
The problem of solving a linear system $Ax=b$ efficiently is a fundamental question in linear algebra, central to both scientific applications and complexity results. Algorithms for linear system solving are generally divided into direct and iterative methods.
The term \emph{direct method} refers to solving $Ax=b$ by applying an (implicit) representation of $A^{-1}$ to $b$ using a decomposition that is exact up to numerical error.
Examples include Gaussian Elimination, Cholesky Factorization, and QR decomposition.
In contrast, iterative methods successively converge to the solution~\cite{S03}. The most basic algorithm among direct methods is Gaussian Elimination, while in the iterative case Conjugate Gradient is most commonly used~\cite{HS52}. A key consideration when developing algorithms for linear systems is the underlying field, as methods for solvers in finite fields, rationals, and reals all differ substantially.

Any algorithm that directly computes $A^{-1}$ leads to a solver for linear equations in $A$.
Strassen formally showed that matrix inversion is equivalent to matrix multiplication in any ring via a divide-and-conquer approach~\cite{S69}. In the RealRAM model, this implies that given a matrix $A \in \mathbb{R}^{n \times n}$ and a vector $b \in \mathbb{R}^n$, the linear system $Ax=b$ is solvable in time $O(n^{\omega})$, where $\omega$ denotes the exponent of matrix multiplication. The constant $\omega$ has a current best bound of $\omega < 2.37286$~\cite{AW21}, which culminates an extensive line of work on fast matrix multiplication based on the analysis of the Coppersmith--Winograd tensor \cite{P80, CW82, S86, W12, LG14, CU13}. Besides square matrix multiplication, rectangular matrix multiplication is central to many problems in algorithm design \cite{LG14}, such as the all-pairs shortest paths problem \cite{Z02} and linear program solvers \cite{CLS21}. On the other hand, iterative algorithms via the Conjugate Gradient or the Lanczos algorithm yield a running time of $O(n \cdot \nnz(A))$ for solving $Ax = b$ in RealRAM \cite{HS52, L50}, where $\nnz(A)$ denotes the number of non-zeroes in the matrix $A$. 

However, running times in the RealRAM model can be misleading. For example, in finite precision arithmetic, even Gaussian Elimination is not stable by default, as first shown by Wilkinson \cite{W61}. Ill-conditioned systems can yield very wrong solutions due to the round-off errors that may arise. But, when restricted to matrices with polynomial condition number, a running time of $O\big(n^{\omega + o(1)}\big)$ can be achieved with guaranteed numerical stability in finite precision arithmetic \cite{DDHK07}. Conjugate Gradient is also not stable in finite precision arithmetic --- as observed in folklore and formally shown in \cite{MMS18}.
If we instead work in the finite field setting, round-off error is no longer a concern, preventing instability. This provides a simpler setting for developing fast linear algebra algorithms, which in turn can shed light into the rational and real cases. 

The first iterative algorithms for the finite field setting were adaptations of previously-known existing methods over the reals, such as the finite field version of the Conjugate Gradient and Lanczos algorithms proposed in~\cite{LO90}. In this case, the motivation for working in finite fields emerged in the area of cryptography; more concretely, in the problems of factoring and the discrete logarithm, which require solving large sparse linear systems over the field $GF(2)$. Nonetheless, this sparked interest for developing iterative algorithms for linear systems directly for finite fields, instead of adapting them from the real setting. A prominent example is Wiedemann's algorithm~\cite{W86}, which yields a probabilistic method for solving linear systems in $O(n\cdot \nnz(A))$ field operations with only a $O(\nnz(A))$ space requirement. Wiedemann's algorithm is based on the observation that, when applying a square matrix repeatedly to a vector, the resulting sequence of vectors is linear recursive. His method then relates the generating polynomial of this sequence with the minimal polynomial of the matrix, which can be computed efficiently over finite fields with the Berlekamp--Massey algorithm \cite{JM89}, among others.

In some contexts, we want to compute $A^{-1}$ instead of merely solving a linear equation $A x = b$.
This may occur, for example, if we need to solve many linear equations in $A$.
Frequently, algorithms for  computing $A^{-1}$ also suggest methods for determining the rank of $A$ (as $A^{-1}$ exists if and only if the matrix has full row and column rank).

In the finite field setting there currently exists a complexity gap between the running time for linear system solving and that of inverse computation of sparse matrices.
Linear systems can be solved in $O(n\cdot \nnz(A))$ time using for example Wiedemann's algorithm~\cite{W86}.
There are also sub-matrix-multiplication-time algorithms for computing the inverse of a sparse matrix \cite{EGGSV06, EGGSV07}, but these are somewhat slower: With $\nnz(A) = \Otil(n)$, the running time is $\Otil(n^{2.28})$.
Until recently, there was no such complexity gap in the real finite precision arithmetic setting, where both running times were $O(n^{\omega})$ \cite{S69}. 
However, \cite{PV21} showed that sparse linear system solving under real finite precision arithmetic can be done faster than matrix-multiplication time, achieving $O(n^{2.331645})$ running time for an $n \times n$ matrix $A$ with $\nnz(A) = \Otil(n)$.

Over rationals, linear systems can be solved exactly using finite precision, making it possible to solve ill-conditioned problems in this setting.
However, the bit complexity of the rational solutions is high, which makes it difficult to obtain fast algorithms.
In particular, if one works with high bit complexity representations of intermediate calculations, this leads to slow implementations of direct methods such as Gaussian Elimination.
Somewhat surprisingly, many of these issues can be resolved by relying on $p$-adic arithmetic for the intermediate calculations, as shown in a pioneering work by \cite{D82}.  
The key idea is to bridge the numerical stability of finite fields with the rationals by means of $p$-adic integers and a ``rational reconstruction'' algorithm, so that one can rely on the finite field numerical stability, and yet recover a rational solution.
Elements of the ring of $p$-adic integers, denoted $\Z_p$ for a prime $p$, are infinite series of powers of~$p$.
Dixon showed that $O(n \log n)$ $p$-adic digits suffice for recovering the exact rational solution via the rational reconstruction algorithm, where each $p$-adic digit can be seen as an element of the finite field $\Z/p\Z$.
For example, by using Dixon's $p$-adic approach, one can merge Wiedemann's algorithm with rational reconstruction to obtain an exact rational solution of an integer linear system~\cite{KS91}.
While Dixon's algorithm has a running time of $O(n^3)$ for linear system solving over the rationals (which is that of Gaussian Elimination, except that in Dixon's algorithm the solution is guaranteed to be exact), Storjohann achieved a running time of $\Otil(n^\omega)$ by leveraging rational reconstruction with a divide-and-conquer method for the $p$-adic expansion \cite{S05}. 

Concurrently, \cite{EGGSV06, EGGSV07} also improved Dixon's algorithm to achieve a running time of $\Otil(n^{2.5})$ for sparse linear system solving over the rationals, and of $\Otil(n^{2.28})$ for sparse matrix inversion over finite fields. In their case, their running time improvement relies on efficient matrix projections and block Krylov methods. Their main ideas on block Krylov methods and structured matrices were recently adapted to the finite precision arithmetic setting to achieve the first sub-matrix-multiplication algorithm for sparse linear systems~\cite{PV21}. In this case, we encounter the reverse situation to the first adaptations of the Conjugate Gradient and Lanczos algorithms to the finite precision real setting, and it is the adaptation of a finite field algorithm to the reals what has achieved a significant running time improvement. This motivates the detailed study of the matrix inversion problem in the finite field setting, given that no sub-matrix-multiplication algorithm for it is known in finite precision real arithmetic.

In this paper, we study the problem of matrix inversion and rank computation of an $n \times n$ matrix $A$ over a finite field, focusing on sparse matrices and certain other classes of structured matrices. In the process, we also study the problems of computing the nullspace when $A$ is singular and the Schur complement of a non-singular principal minor. 

We obtain an expected final running time for all four problems of
\[
\Ohat\big(mn\,\phi(n) + s^{\omega}m + n^{\omega_s} + mn^2\big)
\]
field operations, where $\phi(n)$ denotes the time required to apply $A$ to a vector, $s$ is the blocking factor dividing $n$ and $m$ is its complement (so that $sm = n$, where both $s$ and $m$ are parameters of the algorithm), $\omega$ is the exponent of matrix multiplication \cite{AW21}, and $\omega_s$ is the corresponding exponent for multiplication of an $n\times s$ matrix by an $s\times n$ one. We are using the abbreviation $\omega_s=\omega(\log_n s)$ where $\omega(k)$ is the exponent for multiplication of an $n\times n^k$ matrix by an $n^k \times n$ one, as introduced in the context of fast rectangular matrix multiplication in~\cite{GU18}. The notation $\Ohat(\cdot)$ hides factors $O(n^{o(1)})$. 

In the case where the matrix $A$ is sparse or, more generally, whenever $\phi(n) = \Ohat(n)$, the above running time becomes $\Ohat \big(n^{\omega(k)}\big)$, where $k = \log_n s$ is the only value satisfying $\omega(k) = 3-k$. This corresponds to an expected $O\big(n^{2.2131}\big)$ running time using the current best known bound on $\omega(k)$. Our method relies on the construction of~\cite{EGGSV07} for factoring an arbitrary matrix into block Krylov and block Hankel matrices. We modify their algorithm by inverting the block Hankel matrix explicitly, as opposed to working with its implicit formula. To do so, we employ displacement rank methods combined with fast rectangular matrix multiplication algorithms.


\subsection{Related Work}
Our construction is closely related to the one presented in~\cite{EGGSV07}. They improve Dixon's algorithm for the exact solution of linear systems over the rationals, lowering the running time from $\Tilde{O}(n^3)$ to $\Tilde{O}(n^{2.5})$. Each of the $O(n \log n)$ iterations of Dixon's algorithm requires the application of $A^{-1}$ mod~$p$ to a vector.
Thus, Dixon's running time relies on both inverting the matrix $A$ quickly in $\mathbb{Z}/p\mathbb{Z}$ and then being able to apply it efficiently to a vector. 
The improvements by Eberly et~al.\;\cite{EGGSV07} rely on two central constructions.
First, they introduced efficient block projections which allow for the use of Krylov-type methods without a too high exponentiation of $A$. Scalar block Krylov methods for linear system solving were already in use in the seminal paper of Wiedemann in 1986 for sparse linear systems in finite fields based on the Berlekamp--Massey algorithm~\cite{W86}. 
Still, the $p$-adic version of Wiedemann's algorithm by Kaltofen and Saunders~\cite{KS91} does not improve Dixon's running time, because one needs to apply powers of $A$ up until $A^n$ to a vector at each iteration. This motivates the introduction of blocks to Wiedemann's algorithm, which limits the required powers of $A$ to $A^m$, where $m = n/s$ is the number of blocks. The block version of Wiedemann's algorithm was first proposed by Coppersmith~\cite{C94} through a block generalization of the Berlekamp--Massey algorithm. Shortly after, Kaltofen~\cite{K95} proposed using block Toeplitz systems (which can be solved quickly) instead of the block Berlekamp--Massey algorithm in Coppersmith's algorithm. Eberly et~al.\ introduced efficient block projections $u$ and $v$ in this setting, which enables them to construct $uA^iv$ much faster than in the case of the random block projections in Coppersmith or Kaltofen. The second key ingredient of the Eberly et~al.\ algorithm is the observation that the Gram matrix of the Krylov space matrix is a block Hankel matrix. This leads to a very effective decomposition of $A^{-1}$ mod~$p$ into two Krylov space matrices and the inverse of a block Hankel matrix, which is highly structured. The Krylov space matrices are computed efficiently because the input matrix $A$ is assumed to be sparse, i.e., it has only $\Ohat(n)$ non-zero entries. Thus, it allows for efficient matrix-vector products: computing $A \rightarrow Ab$ only requires $\Ohat(n)$ operations.

While a Hankel matrix appears to lose all of its structure when inverted, Kailath et~al.\ \cite{KKM79} showed how to circumvent this loss. They introduced the notion of \textit{displacement rank}, which consists of applying an invertible linear operator to the Hankel matrix so that its inverse can be expressed as the sum of only two $LU$ products. Bitmead and Anderson~\cite{BA80} used this fact along with FFT convolutions to compute the solution of Toeplitz/Hankel systems in sub-quadratic time. Their algorithm can be extended to the block case by viewing the block matrices as $m \times m$ matrices whose entries are in turn $s \times s$ matrices such that each operation on an $s$-by-$s$ block takes $\Ohat(s^{\omega})$ time. In parallel to the displacement rank methods, Labahn et~al.\;\cite{LCC90} presented a set of formulae for the inverse of a block Hankel or block Toeplitz matrix, which are expressed in terms of certain matrix Pad\'e forms. This is the algorithm that is used by Eberly et~al.\ to invert a block Hankel matrix. However, this approach allows for less generality than the displacement rank method. Beyond structured matrices such as Toeplitz/Hankel and sparse ones, several fast algorithms and hardness results have been developed when considering structured linear systems more broadly, such as those for graph-structured linear systems (e.g., graph Laplacians) \cite{ST14, KZ17, KWZ20}. Laplacian systems have also been recently studied in the finite field setting~\cite{HP20}. 

\subsection{Our Results and Contributions}
In this paper, we improve the current fastest algorithms for sparse matrix inversion, as well as for rank and nullspace computation, over finite fields. In particular, we study the structure of the low displacement rank generators that correspond to the inverse of a block Hankel matrix. Instead of using the Beckermann--Labahn formula as done in \cite{EGGSV07}, we turn to the low displacement rank algorithms for block Toeplitz/Hankel matrices, and observe that the block Hankel inverse can be recovered explicitly from the product of its rectangular generators, which in turn can be done efficiently with the current fast rectangular matrix multiplication algorithms. This yields a final running time for inverting a non-singular $n\times n$ matrix $A$ over a finite field of 
$\Ohat\big(mn\,\phi(n) + s^{\omega}m + n^{\omega_s} + mn^2\big)$.
By using the current best bound on $\omega_s=\omega(\log_n s)$ as given by \cite{GU18} and by optimizing for the block sizes, we obtain an expected final running time of $O\big(n^{2.2131}\big)$ in the case of sparse matrices (Theorem~\ref{thm:inv}). More 
precisely, the running time of our algorithm is equal to $\Ohat \big(n^{\omega(k)}\big)$, where $k$ is the only value satisfying the equation $\omega(k) = 3-k$. Moreover, we obtain the same running time for the computation of the rank and nullspace of sparse matrices over finite fields (Theorem~\ref{thm:rank}), as well as for computing the Schur complement of a non-singular principal minor (Lemma~\ref{lemma:schur}). 


Our algorithm for inverting explicitly a block Hankel matrix (Theorem~\ref{thm:hinv}), which is the building block of our improved running time for sparse matrix inversion in finite fields, extends more generally to other structured matrices. Our construction extends to more general matrix classes where a displacement rank operator exists, and is thus applicable to block Toeplitz-like or Hankel-like matrices (Theorems \ref{thm:expinv} and \ref{blockTH}); i.e., matrices with similar structure with respect to the Toeplitz or Hankel displacement operator, respectively. Our technique is also applicable to other types of displacement operators, such as for block Vandermonde matrices (Theorems~\ref{thm:blockvandermonde}. The use of fast rectangular matrix multiplication combined with rectangular low displacement generators thus provides a new faster scheme for structured block-matrix inversion, which
yields the best current upper bound. We state sufficient conditions that the displacement operator must satisfy in order that our scheme be applicable (Section~\ref{sec:generalstatement}).

\subsection{Applications}
Our results have numerous applications, given that computing the inverse or the rank of a sparse matrix over a finite field is a central problem in linear algebra. We include two applications in topology and algebra. First, we reduce the complexity of Las Vegas type output-sensitive algorithms as in \cite{CK13} for the computation of persistence diagrams in topological data analysis (Theorems~\ref{thm:homology} and \ref{thm:onecomplex}). For this, we rely on the fact that matrices of boundary operators are sparse. Second, motivated by \cite{GJS20}, we provide an improved running time for testing whether an element is a unit in a group ring of a finite metacyclic group~$G$, and if so, computing the $G$-orbit of its inverse (Theorem~\ref{thm:metacyclic}). This is feasible because right-translation matrices in group rings of metacyclic groups are block Toeplitz. 



%

\section{Inversion of Matrices in Finite Fields}\label{sec:inversion}
In this section we present the procedure outlined in Algorithm~\ref{alg:main1}. Section~\ref{sec:displacement} contains the theoretical framework needed for Algorithm~\ref{alg:main2}, which is more general than the former. However, their running times are equivalent and specified in Theorem~\ref{thm:inv}.

\subsection{Preliminaries}
Throughout this paper, $F$ is a finite field, 
$\phi(n)$ is the running time required to apply the input $n \times n$ matrix to a vector, and $s, m$ refer to the blocking factors of a matrix (with $n=sm$). 
For the running times, $\Ohat(\cdot)$ hides factors $O\big(n^{o(1)}\big)$, where $n$ is the dimension of the input matrices (which is generally clear from the context), and $\Otil(\cdot)$ hides logarithmic factors.
Moreover, $\omega < 2.37286$ is the minimum value such that two $n \times n$ matrices can be multiplied using $O\big(n^{\omega + o(1)}\big)$ arithmetic operations \cite{AW21}. Analogously, $\omega(k)$ is the minimum value such that the product between an $n \times n^k$ matrix and an $n^k \times n$ one can be performed using $O\big(n^{\omega(k) + o(1)}\big)$ arithmetic operations \cite{GU18}. We use the abbreviation $\omega_s = \omega(\log_n s)$. Lastly, $\beta > 0.31389$ is the dual exponent of matrix multiplication (Definition~\ref{def:dualexp}), and $\alpha$ refers to the displacement rank of a matrix with respect to some fixed displacement operator (Definition \ref{def:operator}). 

\subsection{Construction}
We begin by recalling the structure presented in Eberly et~al.\;\cite{EGGSV07}. Consider an arbitrary invertible matrix $A$ of size $n \times n$ over a finite field~$F$. We remark that one should first precondition the matrix $A$ as $DAD$, where $D$ denotes the diagonal matrix of indeterminates as defined in \cite[Theorem~2.1]{EGGSV07}, which ensures with high probability the non-singularity of the subsequent Krylov matrices $K_u, K_v$ defined below. However, for notational simplicity, we will keep denoting the matrix by~$A$. We also remark that this preconditioning of $A$ is the only step in the algorithm that causes the final running time to be probabilistic. In particular, all running times given in Section~\ref{sec:displacement} are deterministic.

Let $s \in \mathbb{Z}$ be the blocking size, and let $m = n/s$. Eberly et~al.\ define an efficient block projection in $F^{n \times s}$ as follows. Let
\[
u = \begin{bmatrix}
I_s \\
\vdots \\
I_s
\end{bmatrix}
\]
consist of $m$ copies of $I_s$, the identity matrix of size $s \times s$. We then define the following two Krylov matrices:

\vspace{-0.5cm}

\[
K_u = 
\begin{bmatrix}
 \,\, u \,\,  \,\, Au \,\,  \,\, A^2u \,\,  \,\, \cdots \,\,  \,\, A^{m-1}u \,\,
\end{bmatrix},
\hspace{1cm}
K_v = \begin{bmatrix}
u^T \\[0.1cm]
u^TA \\[0.1cm]
\vdots \\[0.1cm]
u^TA^{m-1} 
\end{bmatrix},
\]
which Eberly et~al.\ show to be non-singular. Both $K_u$ and $K_v$ have size $n \times n$ and $m$ blocks of size $n \times s$ and $s \times n$, respectively. The computation of $K_u$ and $K_v$ requires computing $A^iu$ and $u^TA^i$ for $0 \leq i \leq m-1$. This requires $m-1$ applications of $A$ to $u$, for a total of $O(n\, \phi(n))$ operations. The key insight of Eberly et~al.\ is that $H = K_v A K_u$ is a block Hankel matrix:
\begin{equation}\label{eq:blockh}
    H =  
    \begin{bmatrix}
    u^TAu  & \ldots & u^TA^m u \\
    \vdots & \ddots & \vdots \\
    u^TA^mu & \ldots & u^TA^{2m-1}u \\
    \end{bmatrix} \in F^{n \times n}.
\end{equation}
By the definition of $u$, we can compute $wu$ for any $w \in F^{s \times n}$ with $O(sn)$ operations. Hence, computing each $u^T(A^iu)$ takes $O(sn)$ operations. Finally, we have $0 \leq i \leq 2m-1$ such products, and so the total cost for building the block Hankel matrix $H$ is $O(n^2)$, since $sm = n$.

\subsection{Motivation for Running Time Improvement}
Since $H=K_v A K_u$, computing the inverse $A^{-1}$ amounts to computing $K_u H^{-1} K_v$. There are two ways of proceeding, which fundamentally rely on the question of whether to keep the block Hankel inverse implicit (with, for example, the off-diagonal formula of Beckermann and Labahn~\cite{BL94}; see Theorem~\ref{thm:offdiag}), or to make it explicit before multiplying it with the Krylov matrices. 

After obtaining an efficient representation of $H^{-1}$, Eberly et~al.\ show that we can then compute $H^{-1}M$ for an arbitrary $M \in F^{n \times n}$ in time $\Ohat(s^{\omega}m^2)$. This is the convenient set-up for solving a linear system in a Dixon-like scheme, since we need to be able to apply a vector efficiently to $H^{-1}$ at each iteration. We propose a different scheme for explicitly inverting $A^{-1}$. The Eberly et~al.\ construction does not take advantage of the Krylov structure of $K_u$ when computing $H^{-1}K_u$, and instead treats $K_u$ as an arbitrary matrix. However, multiplying an arbitrary matrix with $K_u$ or $K_v$ takes only $O(mn^2)$. Thus, we propose the following alternative construction:
\begin{enumerate}
    \item After inverting $H^{-1}$ efficiently and obtaining an implicit formula for the inverse, we recover $H^{-1}$ \textit{explicitly} with fast rectangular matrix multiplication.
    \item Next we treat $H^{-1}$ as an arbitrary matrix and compute $H^{-1}K_u$ by exploiting the Krylov structure of $K_u$. Finally, we compute $K_v \cdot (H^{-1}K_u)$ by using the Krylov structure of $K_v$.
\end{enumerate}

\subsection{Inverting a Block Hankel Matrix Explicitly}\label{sec:hinv}
First, we need to compute the inverse of the block Hankel matrix $H^{-1}$. There are several efficient algorithms to invert (block) Toeplitz/Hankel matrices, which generally fall into two categories: either they use the low displacement structure as introduced by Kailath et~al.\;\cite{KKM79} or they build on the inverse formulae of Gohberg--Semencul~\cite{GS72} and Trench~\cite{T64}. In this section we focus on the second kind, since it is the method followed by Eberly et~al. However, we will then argue that the displacement rank method is much more general and applicable in other settings, so in Section~\ref{sec:displacement} we will turn to the low displacement rank methods. Building on the Gohberg--Semencul, Heining, and Krupnik formulae~\cite{GS72, GK72, GH74}, Labahn et~al.\;\cite{LCC90} generalized their methods to block Hankel matrices and presented a new set of formulae for the inverse of block Hankel/Toeplitz matrices which only requires their non-singularity. They did so by representing $H^{-1}$ with matrix Pad\'e forms, as shown in~\cite{LC89}.

\newpage

\begin{theorem}[{\cite[Theorem~3.1]{LCC90}}]\label{thm:offdiag}
Given a block Hankel matrix $H$ with blocks of size $s \times s$ and $m$ blocks in each row/column (where $sm=n$), 
the inverse $H^{-1}$ can be expressed as
\[
H^{-1} = \begin{bmatrix}
    v_{m-1}  & \ldots & v_1 & v_0 \\
    \vdots & \reflectbox{$\ddots$} & \reflectbox{$\ddots$} &  \\
    v_1 &  \reflectbox{$\ddots$} &  & \\
    v_0 &  &  & \\
    \end{bmatrix} \cdot 
    \begin{bmatrix}
    q^{*}_{m-1} & \cdots & q_0^{*} \\
    & \ddots & \vdots \\
    & & q_{m-1}^{*}
    \end{bmatrix}
- \begin{bmatrix}
    q_{m-2}  & \ldots & q_0 & 0 \\
    \vdots & \reflectbox{$\ddots$} & \reflectbox{$\ddots$} &  \\
    q_0 &  \reflectbox{$\ddots$} &  & \\
    0 &  &  & \\
    \end{bmatrix} \cdot 
    \begin{bmatrix}
    v^{*}_{m} & \cdots & v_1^{*} \\
    & \ddots & \vdots \\
    & & v_{m}^{*}
    \end{bmatrix},
\]
where $v_i$, $v^{*}_i$, $q_i$ and $q_i^{*}$ are $s\times s$ matrices.
\end{theorem}

Let us denote the four matrices in Theorem~\ref{thm:offdiag} as
\[
H^{-1} = V Q^{*} - Q V^{*},
\]
where $V$ and $Q$ are anti-triangular block Hankel matrices and $V^{*}$ and $Q^{*}$ are triangular block Toeplitz matrices. As noted by Eberly et~al., by using the fast algorithms for Pad\'e formulations from~\cite{GJV03}, the matrices $V$, $Q$, $V^{*}$, $Q^{*}$, and thus the implicit representation of $H^{-1}$ in Theorem~\ref{thm:offdiag} (which is also known as the \textit{off-diagonal inverse formula}), can be computed with $\Ohat(s^{\omega}m)$ operations in~$F$. The question is then: what is the most efficient way to recover $H^{-1}$ \textit{explicitly} from its implicit representation above? 

By using fast algorithms for matrix polynomials~\cite{CK91}, we can compute the product $H^{-1}M$ for an arbitrary $M \in F^{n \times n}$ in time $\Ohat(s^{\omega}m^2)$. Thus, by setting $M = I_n$, or just by treating $Q^{*}$ and $V^{*}$ as arbitrary matrices, we can recover $H^{-1}$ explicitly in $\Ohat(s^{\omega}m^2)$. We would like to do better, given that $Q^{*}$ and $V^{*}$ have a very particular structure and are not arbitrary matrices. 

To obtain a better upper bound for the explicit recovery of a block Hankel inverse matrix, we will instead use fast rectangular matrix multiplication. Given that $V, Q, V^{*}, Q^{*}$ are (anti-) triangular block Hankel and Toeplitz matrices, we can associate rectangular matrices of sizes $n \times s$ and $s \times n$ to them, which consist of the $m$ non-repeated blocks. In other words, we define:
\begin{equation}\label{eq:vbar}
    \overline{V} = 
    \begin{bmatrix}
    v_{m-1} \\
    \vdots \\
    v_1 \\
    v_0
    \end{bmatrix}, 
    \quad
    \overline{Q} = 
    \begin{bmatrix}
    q_{m-2} \\
    \vdots \\
    q_0 \\
    0
    \end{bmatrix} \in F^{n \times s},
\end{equation}
\begin{equation}\label{eq:qbar}
    \overline{Q*} = 
    \begin{bmatrix}
     \,\, q^*_{m-1} \,\, | \,\, \cdots \,\, | \,\, q^*_0 
    \end{bmatrix},
    \quad
    \overline{V^{*}} = 
    \begin{bmatrix}
     \,\, v^{*}_{m} \,\, | \,\, \cdots \,\, | \,\, v_1^{*} \,\,
    \end{bmatrix} \in F^{s \times n}
\end{equation}
from $V, Q^{*}, Q, V^{*}$. Our key insight is that to compute $VQ^{*}$ we can instead perform the rectangular product $\overline{V}\, \overline{Q^{*}}$ and then recover $VQ^{*}$ from $\overline{V}\,\overline{Q^{*}}$ in $O(n^2)$ time by adding through each anti-diagonal. This is because we have the following correspondence between $VQ^{*}$ and~$\overline{V}\,\overline{Q^{*}}$:
\begin{equation}\label{eq:recovery}
    [VQ^{*}]_{i, j} = \sum_{1 \leq k\leq \min\{j,\,m-i+1\}} [\overline{V}\, \overline{Q^{*}}]_{i+k-1,\, j-k+1}
\end{equation}
where $[VQ^{*}]_{i, j}$ denotes the $s \times s$ block of $VQ^{*}$ located at row $i$ and column~$j$.

To put it more visually, after performing the rectangular product $\overline{V}\, \overline{Q^{*}}$, we obtain the $n \times n$ matrix
\[
\overline{V}\, \overline{Q^{*}} = 
\begin{bmatrix}
v_{m-1}q^{*}_{m-1} & v_{m-1}q^{*}_{m-2} & \cdots & v_{m-1}q^{*}_{0} \\[0.3cm]
v_{m-2}q^{*}_{m-1} & v_{m-2}q^{*}_{m-2} & \cdots & v_{m-2}q^{*}_{0} \\[0.3cm]
\vdots & \vdots & \ddots & \vdots \\[0.2cm]
v_0q^{*}_{m-1} & v_0q^{*}_{m-2} & \cdots & v_0q^{*}_{0}
\end{bmatrix}.
\]
We can then  build $VQ^{*}$ from $\overline{V}\,\overline{Q^{*}}$ by adding through each anti-diagonal one block at a time, for a total of $O(n^2)$, since 
\[
VQ^{*}=
\begin{bmatrix}
v_{m-1}q^{*}_{m-1} & \quad v_{m-1}q^{*}_{m-2} + v_{m-2}q^{*}_{m-1} & \quad \cdots &  \quad \sum_{k=1}^m v_{m-k} q^{*}_{k-1} \\[0.3cm]
v_{m-2}q^{*}_{m-1} & \quad v_{m-2}q^{*}_{m-2} + v_{m-3}q^{*}_{m-1} 
& \quad \cdots & \quad \sum_{k=2}^{m} v_{m-k} q^{*}_{k-2}\\[0.3cm]
v_{m-3}q^{*}_{m-1} & \quad v_{m-3}q^{*}_{m-2}+v_{m-4}q^{*}_{m-1} &  \quad \cdots & \quad\sum_{k=3}^{m} v_{m-k} q^{*}_{k-3}\\[0.3cm]
\vdots & \quad \vdots  & \quad \ddots & \quad \vdots \\[0.2cm]
v_0 q^{*}_{m-1} & \quad v_0 q^{*}_{m-2} & \quad \cdots & \quad v_0 q^{*}_{0}
\end{bmatrix}.
\]
We thus obtain the following cost:
\begin{theorem}\label{thm:hinv}
For a block Hankel matrix $H \in F^{n \times n}$ with blocking size $s = n/m$, computing $H^{-1}$ explicitly requires $\Ohat \big(n^{\omega_s}\big)$ field operations, which corresponds to the running time required to multiply an $n\times s$ matrix with an $s\times n$ matrix.
\end{theorem}

\begin{proof}
Computing the product $\overline{V}\,\overline{Q^{*}}$ costs $n^{\omega_s}$.
Then, we recover $VQ^{*}$ from $\overline{V}\,\overline{Q^{*}}$  
using Equation~\eqref{eq:recovery}.
This recovery 
only needs reading through the entries of $\overline{V}\,\overline{Q^{*}}$, which costs $O(n^2)$. The 
same reasoning applies to the product $QV^{*}$. 
\end{proof}

Once $H^{-1}$ has been made explicit, we need to multiply it on both sides by the Krylov matrices $K_u$ and $K_v$ to obtain $A^{-1}$:
\[
A^{-1} = K_u H^{-1} K_v.
\]
Not only can we multiply $K_v$ efficiently with 
any $M$ from the right ($K_vM$), as shown by Eberly et~al., but we can also do it from the left ($MK_v$). Hence, after obtaining $H^{-1}$ explicitly, we perform the product $H^{-1} K_v$ as follows, by effectively treating $H^{-1}$ as an arbitrary matrix $M$ now that it has been made explicit. Split $H^{-1}$ into $m$ blocks of $s$ consecutive columns $H^{-1}_i$ for $0 \leq i \leq m-1$. Using Horner's scheme to apply $K_u$ to each block and summing the results, we then obtain
\begin{align*}
H^{-1}K_u & = \sum_{i=0}^{m-1} H^{-1}_0 u^T A^i \\ & = (\cdots(H^{-1}_{m-1}u^TA + H^{-1}_{m-2}u^T)A + 
H^{-1}_{m-3}u^T)A + \cdots  + H^{-1}_1u^T)A + H^{-1}_0u^T.
\numberthis \label{eq:horner1}
\end{align*}
By the special structure of $u^T$, we can compute each $H^{-1}_i u^T$ in $O(n^2)$, yielding a total of $O(mn^2)$. Then we multiply each $H^{-1}_iu^T$ by $A$, and since there are a total of $m$ such products, the final running time of computing $MK_u$ is again $O(mn\,\phi(n) + mn^2)$.

Finally, we perform the product $K_v \cdot (H^{-1}K_u)$ by now treating $H^{-1}K_u$ as an arbitrary matrix. We obtain the same running time as for $H^{-1}K_u$ by performing a similar construction. Split $H^{-1}K_u$ into $m$ blocks of $s$ consecutive rows $(H^{-1}K_u)_i$, for $0 \leq i \leq m-1$. Now we 
obtain
\begin{align*}
K_v \cdot ( & H^{-1} K_u) = \sum_{i=0}^{m-1} 
A^i\,u(H^{-1}K_u)_i 
\\ & = u(H^{-1}K_u)_0 + A[u(H^{-1}K_u)_1 + 
A[u(H^{-1}K_u)_2 + \cdots + 
A\,u(H^{-1}K_u)_{m-1}] \cdots ].
\numberthis \label{eq:horner2}
\end{align*}
By the same 
argument as before, the final running time for this product is $O(mn\,\phi(n) + mn^2)$.

The total cost for building $A^{-1}$ is then
\[\,\,\,
\Ohat\big(mn\,\phi(n) + s^{\omega}m + n^{\omega_s} + mn^2\big).
\]
In the sparse case where $\phi(n) = \Ohat(n)$, the above becomes
$\Ohat(s^{\omega}m + n^{\omega_s} + mn^2)$.

For the proof of Theorem~\ref{thm:inv} below, we need to quote the following fact:
\begin{lemma}[{\cite[Eq.\,2.6]{HP98}}]\label{lemma:pan}
Multiplying an $n\times s$ matrix by an $s\times n$ matrix can be done with the same number of arithmetic operations as
multiplying an $s\times n$ matrix by an $n\times n$ matrix.
\end{lemma}

\begin{theorem}\label{thm:inv}
For a non-singular matrix $A \in F^{n \times n}$ where $F$ is a finite field, the inverse $A^{-1}$ can be computed
in expected time
\[
\Ohat(mn\,\phi(n)+s^{\omega}m + n^{\omega_s} + mn^2),
\]
where $sm=n$, by calling the procedure \textsc{MatrixInv}$(A)$ (this applies to both versions $1$ and $2$; see Algorithm~{\rm\ref{alg:main1}} and Algorithm~{\rm\ref{alg:main2})}.
If~$\phi(n) = \Ohat(n)$, e.g., if $A$ is sparse, then the inverse $A^{-1}$ can be computed in expected time
\[
\Ohat \big(n^{\omega(k)}\big),
\]
where $k = \log_n s$ is the only value that satisfies $\omega(k) = 3-k$. With the current values of rectangular matrix multiplication, this corresponds to $O\big(n^{2.2131}\big)$ arithmetic operations.
\end{theorem}

\begin{algorithm}\caption{Inverting an arbitrary matrix over a finite field with the off-diagonal formula}\label{alg:main1}
\begin{algorithmic}[1]
\Procedure{\textsc{MatrixInv1}}{$A$} \Comment{Theorem~\ref{thm:inv}}
    \State Fix $s$ and $m$ blocking factors such that
    $n=sm$. \Comment{
      Values $s, m \leftarrow$ $n^{0.7869}, n^{0.2131}$}
    \\  \Comment{are optimal at the 
      current rectangular matrix multiplication time.}
    \State $u \leftarrow [I_s \cdots I_s]^T$
    \State $K_v \leftarrow [u^T  \,\,\, u^TA  \,\,\, u^TA^2  \,\,\, \cdots  \,\,\, u^TA^{m-1}]^T$
    \State $H \leftarrow K_v A K_u$ is block Hankel. \Comment{Equation \eqref{eq:blockh}, \cite{EGGSV07}}
    \State $H^{-1} \leftarrow VQ^{*} - QV^{*}$ \Comment{Theorem \ref{thm:offdiag}, \cite{LCC90} 
    }
    \State $\overline{V}, \overline{Q^{*}}, \overline{Q}, \overline{V^{*}} \leftarrow$ as defined in Equations \eqref{eq:vbar}, \eqref{eq:qbar}.
    \State Perform the two rectangular products $\overline{V} \overline{Q^{*}}$ and $\overline{Q} \overline{V^{*}}$. \Comment{Fast algorithm by \cite{GU18}}
    \State Recover $VQ^{*}$ and $QV^{*}$ recursively. \Comment{Equation \eqref{eq:recovery}}
    \State Compute $A^{-1} \leftarrow K_u H^{-1} K_v$. \Comment{Equations \eqref{eq:horner1}, \eqref{eq:horner2}}
    \State \Return $A^{-1}$.
\EndProcedure
\end{algorithmic}
\end{algorithm}

\begin{algorithm}\caption{Inverting an arbitrary matrix over a finite field with displacement rank operators}\label{alg:main2}
\begin{algorithmic}[1]
\Procedure{\textsc{MatrixInv2}}{$A$} \Comment{Theorem~\ref{thm:inv}}
    \State Fix $s$ and $m$ blocking factors such that
    $n=sm$. \Comment{
      Values $s, m \leftarrow$ $n^{0.7869}, n^{0.2131}$}
    \\  \Comment{are optimal at the 
      current rectangular matrix multiplication time.}
    \State $u \leftarrow [I_s \cdots I_s]^T$
    \State $K_v \leftarrow [u^T  \,\,\, u^TA  \,\,\, u^TA^2  \,\,\, \cdots  \,\,\, u^TA^{m-1}]^T$
    \State $H \leftarrow K_v A K_u$ is block Hankel. \Comment{Equation \eqref{eq:blockh}, \cite{EGGSV07}}
    \State Compute $X_i, Y_i$ 
    so that $\Delta_{Z_0^T\!,\, Z_0}(H^{-1}) = \sum_{i=1}^{\alpha} X_iY_i^T$. \Comment{Def.\;\ref{def:operator}, \cite{BA80}
    (Thm.\;\ref{thm:ba80})}
    \State Perform the $\alpha$ rectangular products $X_iY_i^T$. \Comment{Fast algorithm by \cite{GU18}}
    \State Recover $H^{-1}$ recursively. \Comment{Equation \eqref{eq:hankel}}
    \State Compute $A^{-1} \leftarrow K_u H^{-1} K_v$. \Comment{Equations \eqref{eq:horner1}, \eqref{eq:horner2}}
    \State \Return $A^{-1}$.
\EndProcedure
\end{algorithmic}
\end{algorithm}


\begin{proof}
We first note that $s^{\omega}m \le n^{\omega_s}\le s^{\omega}m^2$. The second inequality follows from the definition of $\omega_s$, since 
multiplying an $n\times s$ matrix by an $s\times n$ one can be done with $m^2$ multiplications of $s\times s$ matrices. 
Since it is possible to multiply two $n\times n$ matrices by multiplying $m$ times an $s\times n$ matrix by an $n\times n$ matrix by Lemma~\ref{lemma:pan}, we conclude that $mn^{\omega_s}\ge n^{\omega}$. Consequently,
\[
n^{\omega_s}\ge m^{-1}n^{\omega}=m^{-1}s^{\omega}m^{\omega}=s^{\omega}m^{\omega-1}\ge s^{\omega}m,
\]
which yields the first inequality.

Since $s^{\omega}m \le n^{\omega_s}$, we need to pick optimal blocking factors $s$ and $m$ with $sm = n$ such that $\max\{n^{\omega_s}, mn^2\}$ is minimized. Note that 
\[
mn^2=(ns^{-1})n^2=n^{3-\log_n s}
\]
is a decreasing function of $s$ while $n^{\omega_s}$ is an increasing function of $s$ depicted in Figure~\ref{fig:plot} in \cite{GU18} and in Figure~\ref{fig:plot} below. Hence the optimal value of $s$ is achieved at the crossing point between the graphs of $n^{\omega_s}$ and $mn^2$, which occurs at the only value of $k=\log_n s$ that satisfies 
\[
\omega(k)=3-k.
\]
Using the data given in Table~3 of \cite{GU18}, we find by interpolation the solution $k=0.7869$, corresponding to $s = n^{0.78668}$ and $\omega_s=\omega(0.78668) = 2.21312$. 
This yields a running time bound for inverting a sparse $n\times n$ matrix over a finite field  of
$O\big(n^{2.2131}\big)$.
\end{proof}


\bigskip

\begin{figure}[t]
\centering
\begin{tikzpicture}
\begin{axis}[
    legend cell align=left,
    width=9.5cm, 
    height=6.5cm,
    xmin=0, xmax=1.1,
    ymin=1, ymax=3,
    thick,
    scale only axis,
    xmajorgrids,
    ymajorgrids,
    xtick={0,0.31389,0.7869,1},
    xticklabels={0,0.3139,0.7869,1},
    ytick={1,2,2.2131,2.3729,3},
    yticklabels={1,2,2.2131,2.3729,3},
    xlabel={$k$},
legend style={at={(0.772,0.2)},
anchor=west,legend columns=1},
legend style={font=\small},
legend entries={$mn^2$, $s^{\omega}m^2$, $n^{\omega_s}$, $s^{\omega}m$}
]

\addplot[densely dashed,black] {3-x}; 
\addplot[dotted,black] {2+0.3729*x}; 
\addplot [solid, every mark/.append style={solid, fill=gray}] coordinates {
(0,2)
(0.31477,2.000001)
(0.31508,2.000002)
(0.31532,2.000003)
(0.31552,2.000004)
(0.32,2.000064)
(0.33,2.000448)
(0.34,2.001118)
(0.35,2.001957)
(0.36,2.0031) 
(0.37,2.0045) 
(0.375,2.0053)
(0.3825,2.0067)
(0.4,2.010314)
(0.4125,2.0135)
(0.425,2.0169)
(0.4375,2.0207)
(0.45,2.024801)
(0.47,2.0321)
(0.5,2.044183)
(0.5286,2.057085)
(0.55,2.067488)
(0.6,2.093981)
(0.65,2.123097)
(0.7,2.154399)
(0.75,2.187543)
(0.8,2.222256)
(0.85,2.258317)
(0.9,2.295544) 
(0.95,2.333789)
(1,2.372927)
};
\addplot[densely dotted,black] {1+1.3729*x}; 
\addplot [only marks, every mark/.append style={solid, fill=white}, mark=*] coordinates {
(0.31389,2)
(1,2.372927)
};
\addplot [only marks, every mark/.append style={solid}, mark=*] coordinates {
(0.7869,2.2131)
};
\end{axis}
\end{tikzpicture}
\caption{\label{fig1} Graphic visualization of the minimization process that yields our running time. As functions of $k=\log_n s$, we have $n^{\omega_s}=n^{\omega(k)}$, $mn^2=n^{3-k}$, $s^{\omega}m=n^{1+(\omega-1)k}$, and $s^{\omega}m^2=n^{2+(\omega-2)k}$. The curves plot the respective exponents for comparison. The value $0.3139$ corresponds to the current best bound on the dual exponent of matrix multiplication \cite{GU18}.}\label{fig:plot}
\end{figure}
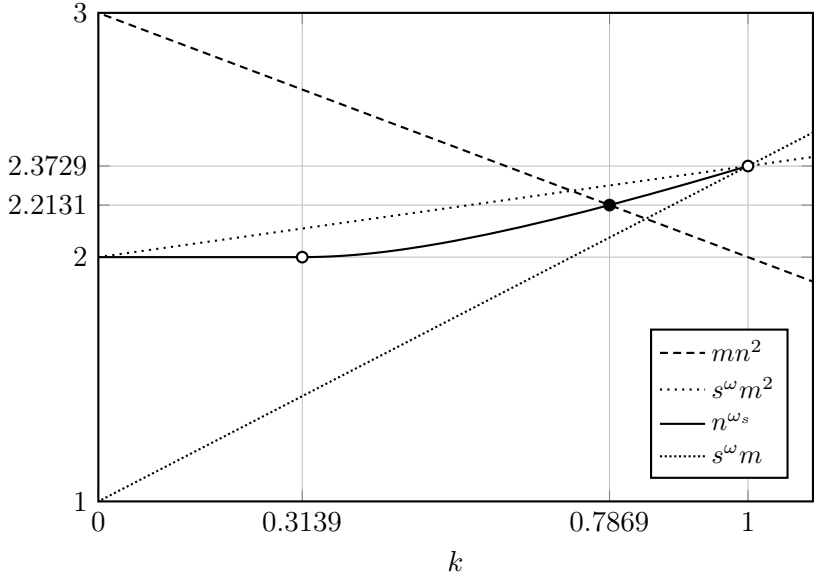




\section{Generalization to Structured Matrices}\label{sec:displacement}
The idea of using rectangular matrix multiplication on the low rank generators of a structured matrix extends to more general settings. Since the off-diagonal inverse formula from the previous section relates only to Hankel matrices, we switch to low displacement rank methods for matrix inversion. First, these are simpler algorithms than the Pad\'e-based ones, and they extend more naturally to the block setting. Second, they allow us to obtain running times not only for Toeplitz and Hankel matrices, but for Toeplitz-like and Hankel-like matrices, as well as other types of structured matrices. 

The notion of low displacement rank was first introduced by Kailath et~al.\;\cite{KKM79}, and referred only to Toeplitz matrices:
\begin{definition}[\cite{KKM79}]\label{def:displrank}
The $(+)$-\textit{displacement rank} of a matrix $M$ is the smallest integer $\alpha_{+}(M)$ such that we can write
\[
M = \sum_{i=1}^{\alpha_{+}(M)} L_i U_i
\]
for some lower-triangular Toeplitz matrices $\{L_i\}$ and upper-triangular Toeplitz matrices $\{U_i\}$.
\end{definition}

The $(-)$-displacement rank is defined similarly, by replacing $L_iU_i$ with $U_iL_i$.
The key theorem in displacement rank methods states the following:
\begin{theorem}[{\cite[Theorem 1]{KKM79}}]\label{thm:kkm79}
The $(\pm)$-displacement rank of a matrix is equal to the $(\mp)$-displacement rank of its inverse, i.e., for all non-singular matrices $M$,
\[
\alpha_{+}(M) = \alpha_{-}(M^{-1}) \quad \text{and} \quad \alpha_{-}(M) = \alpha_{+}(M^{-1}).
\]
\end{theorem}

Soon after, it became evident that the notion of displacement rank can be applied to other types of structured matrices beyond Toeplitz. Following the notation of~\cite{BJS07}, we can then refer to the notion of displacement rank in greater generality, which extends Definition~\ref{def:displrank} to other kinds of operators.

\begin{definition}\label{def:operator}
    Given a matrix $A \in F^{n \times n}$, let $\Delta_{P,Q}$ denote the \textit{displacement operator} of $A$, for $P, Q \in F^{n \times n}$, 
    which takes the form
    \[
    \Delta_{P,Q}(A)=A - PAQ.
    \]
    Two matrices 
    $X, Y \in F^{n \times \alpha}$
    are called \textit{generators of length~$\alpha$} for $A$ if $\Delta_{P, Q}(A) = XY^T$.  In the block case, generators are rectangular matrices of size $n \times \alpha s$. For any matrix $A$ and its associated operator $\Delta_{P, Q}$, the value $\alpha = \textrm{rank}(\Delta_{P, Q}(A))$ is called the \textit{displacement rank} of $A$.
\end{definition}

In this context, we are always interested in the case where $\alpha$ is small relative to $n$, i.e., when $\alpha = o(n)$, and then we say that the matrix $A$ is \textit{$\Delta_{P, Q}$\nobreakdash-like}, or that it has a structure of type $\Delta_{P, Q}$. This is what we mean throughout this section by \textit{Toeplitz-like} or \textit{Hankel-like} matrices. Thus, this notion extends well-beyond the definitions in~\cite{KKM79} for Toeplitz matrices, not only because we allow other types of structured matrices, but also because $\alpha$ can be any constant other than 2. Remarkably, the proof given in \cite{KKM79} for Theorem~\ref{thm:kkm79} does not depend on the definition of the operator, and only requires some general rank properties to hold. In contrast, the off-diagonal formula of Beckermann and Labahn does not allow for such generalizations.
Moreover, the displacement rank algorithms are more readily generalizable to the block setting, which we require.

Closely related to the rectangular matrices in the off-diagonal formula of Beckermann and Labahn in the previous section, these generators are also compact data structures representing~$A$. 
When $A$ has low displacement rank with respecto to $\Delta_{P, Q}$, we can represent $\Delta_{P, Q}(A)$ with two matrices that have size only $n \times \alpha$, hence using a total space of $2 n \alpha$ instead of $n^2$.
We always choose $P$ and $Q$ such that $\Delta_{P, Q}$ is an invertible linear operator.
Hence we can also recover $A$ from the compact representation of $\Delta_{P, Q}(A)$.

For $f \in \Z$, define the circulant matrix:
\begin{equation}
Z_f = \begin{bmatrix}
0 & & & f \\
1 & \ddots & & \\
& \ddots & \ddots & \\
& & 1 & 0
\end{bmatrix}.
\end{equation}
Then, in the case of a Toeplitz matrix $T$, the generators $P, Q$ correspond to $P = Z_0$ and $Q= Z_0^T$. In the case of a Hankel matrix, the matrices $P, Q$ correspond to $P = Z_0^T$ and $Q= Z_0$. It is clear that Toeplitz matrices $T$ and Hankel matrices $H$ have displacement rank~$2$, but the key insight is that now both $\Delta_{P, Q}(T^{-1})$ and $\Delta_{P, Q}(H^{-1})$ have rank~$2$ as well. Displacement operators thus yield compressed bilinear expressions for structured matrices.

\subsection{Explicit Inversion of Low Displacement Rank Matrices}
The notion of displacement rank extends naturally to blocked matrices. We first consider the case of explicitly inverting a block Toeplitz/Hankel-like matrix, using the notation in~\cite{PV21}. Throughout this section we will use the notation for Toeplitz-like matrices (including Algorithm~\ref{alg:blockth}), and then show how the analogous results follow for Hankel-like matrices. The following theorem applies when using the Toeplitz operator with $P= Z_0$, $Q = Z_0^T$. Let $A \in F^{n \times n}$ be a block matrix with block size $s \times s$ and $m \times m$ blocks.

\begin{theorem}[\cite{KKM79, BA80}]
\label{3.4}
Given generators $X_i, Y_i$ of size $n \times s$ and displacement rank~$\alpha$, the equation
\[
A - Z_0\,A\,Z_0^T= \sum_{i=1}^{\alpha} X_i Y_i^T
\]
has the unique solution
\begin{equation}
    A = \sum_{i=1}^{\alpha} L(X_i)U(Y_i),
\end{equation}
where $L(X_i)$ is a lower-triangular Toeplitz matrix whose first column is~$X_i$, and $U(Y_i) = L(Y_i)^T$.
\end{theorem}

Here we use the following correspondence between rectangular matrices $W \in F^{n \times s}$ and $n \times n$ lower (or upper) Toeplitz triangular matrices:
\[
W = 
\begin{bmatrix}
w_{1} \\
w_{2} \\
w_{3} \\
\vdots \\
w_{m}
\end{bmatrix} \in F^{n \times s},
\hspace{0.3cm}
L(W) = 
\begin{bmatrix}
w_{1} & 0 & 0 & \ldots & 0 \\
w_{2} & w_{1} & 0 & \ldots & 0 \\
w_{3} & w_{2} & w_{1} & \ldots & 0 \\
\vdots & \vdots & \vdots & \ddots & \vdots \\
w_{m} & w_{m-1} & w_{m-2} & \ldots & w_{1}
\end{bmatrix} \in F^{n \times n}, \hspace{0.3cm}
U(W)=L(W)^T.
\]

The functional equation $A - Z_0\,A\,Z_0^T= \sum_{i=1}^{\alpha} X_iY_i^T$ is coherent with the definition of the generators $X$ and $Y$ of length $\alpha$ from the previous section. We can either express $\Delta_{Z_0,\, Z_0^T} (A)$ as $XY^T$, where $X$ and $Y$ have size $n \times \alpha s$, or we can express $\Delta_{Z_0,\, Z_0^T} (A)$ as $\sum_{i=1}^{\alpha} X_i Y_i^T$, where each $X_i$ or $Y_i$ has size $n \times s$. 

The algorithm presented in Bitmead and Anderson~\cite{BA80} uses Theorem~\ref{thm:kkm79} applied to the inverse $T^{-1}$ of a Toeplitz matrix:

\begin{theorem}[{\cite[Theorem 2]{BA80}}]\label{thm:ba80}
Given as input a blocked Toeplitz matrix $T$ with blocking size $s = n/m$, the rectangular generators $X_i, Y_i$ of $\Delta_{Z_0,\, Z_0^T}(T^{-1})$ can be found in time can be found in time $\Ohat(s^{\omega}m \log^2 m)$.
\end{theorem}

\begin{remark}\label{rem:ba80}
While the \cite{BA80} algorithm refers specifically to Toeplitz matrices, it is easily extendable to Hankel matrices with the same runtime, as formally discussed in \cite{P90, P01}.
\end{remark}

Bostan, Jeannerod, and Schost \cite{BJS07} later generalized the \cite{BA80} algorithm to arbitrary matrices with large displacement rank $\alpha$, showing that the rectangular generators of the inverses of Toeplitz, Hankel, Vandermonde and Cauchy-like matrices can be found in time $\Ohat(\alpha^{\omega-1}s^{\omega}m \log^2 m)$. The exponent of $\alpha$ is an improvement over Kaltofen's algorithm, who had previously shown a running time of $\Ohat(\alpha^{2}s^{\omega}m \log^2 m)$ for Toeplitz-like matrices \cite{K94}. 

\begin{theorem}[{\cite{BJS07}}]\label{thm:bjs}
Given as input a blocked Toeplitz/Hankel-like matrix $A$ with blocking size $s = n/m$ and displacement rank $\alpha$, the rectangular generators $X_i, Y_i$ of $\Delta_{Z_0,\, Z_0^T}(A^{-1})$ can be found in time $\Ohat(\alpha^{\omega -1}s^{\omega}m \log^2 m)$.
\end{theorem}

All of the \cite{BA80, BJS07, K94} schemes are very efficient to then apply a vector to $T^{-1}$, since the rectangular blocked structure of $X_i$ and $Y_i$ allows for the use of FFT to perform convolutions between the vector and the blocks of $X_i$ and $Y_i$. However, in some applications, as shown in Sections \ref{sec:inversion} and \ref{sec:rank} in this paper, it is convenient to be able to retrieve the block Toeplitz matrix explicitly from its rectangular generators, such as in the case of sparse matrix inversion. In this case, none of the methods for structured matrix inversion have addressed the upper bound for \textit{explicit} inversion, and the running times are always given assuming that the representation of the structured matrices remains implicit. There is no extensive analysis for the decompression stage; i.e., when we want to recover $A^{-1}$ from the rectangular generators of $\Delta_{P, Q}(A^{-1})$. In this setting, the only known way to recover $A^{-1}$ explicitly is to apply the $n/s = m$ canonical block vectors. This requires $m$ convolutions with $A^{-1}$, which yields a total running time of $\Ohat(s^{\omega}m^2 \log^2 m)$. 

Rather, we can apply our scheme to recover 
\begin{equation}\label{eq:toep1}
    A^{-1} = \sum_{i=1}^{\alpha} L(X_i)U(Y_i)
\end{equation}
explicitly, where $\alpha$ is the displacement rank, as usual. We apply a similar approach to what we did in Section~\ref{sec:inversion} for the Beckermann--Labahn formula. Note that the off-diagonal formula for $H^{-1}$ of Section~\ref{sec:inversion} is a special case of displacement in the Hankel case, with $P = Z_0^T$ and $Q = Z_0$.

We first multiply the rectangular matrices $X_i, Y_i \in F^{n \times s}$ to obtain $X_i Y_i^T$ in time $\Ohat \big(n^{\omega_s}\big)$. Using fast rectangular matrix multiplication methods, this can be done better than in time $\Ohat \big(s^{\omega}m^2\big)$. To then recover $L(X_i)U(Y_i)$ from $X_iY_i^T$, we can just read along each diagonal (thus performing only $O(m^2)$ sums of $s \times s$ matrices), since
\begin{equation}\label{eq:lu}
    [L(X_i)U(Y_i)]_{j, k} = \sum_{1 \leq \ell \leq \min\{j, k\}} [X_i Y_i^T]_{j+1-\ell,\, k+1-\ell}.
\end{equation}

Because we need to repeat this procedure $\alpha$ times, we obtain
the following final running time.
\begin{theorem}\label{thm:expinv}
Given a blocked matrix $A \in F^{n \times n}$ with displacement rank $\alpha$ with respect to the Toeplitz/Hankel operator
and blocking size $s = n/m$, 
the inverse $A^{-1}$ can be recovered explicitly from its low rank representation in time 
\[
\Ohat\big(\alpha \big(n^{\omega_s} + n^2\big)\big) = \Ohat\big(\alpha n^{\omega_s}\big).
\]
Hence, $A^{-1}$ can be computed explicitly by calling the procedure \textsc{BlockStructInv}$(A, s, m)$ (see Algorithm~\rm{\ref{alg:blockth}}) in total time $\Ohat\big(\alpha^{\omega-1} s^{\omega}m \log m + \alpha n^{\omega_s}\big) = \Ohat\big(\alpha^{\omega-1}n^{\omega_s}\big)$.
\end{theorem}

In particular, the running time given in Theorem~\ref{thm:expinv} is an improvement over $\Ohat(s^{\omega}m^2)$ for all structured 
matrices with respect to the Toeplitz/Hankel displacement operator, i.e., for those matrices such that
$\alpha = o(n^{1/(\omega-1)})$ with $\alpha = \textrm{rank}(\Delta_{Z_0,\, Z_0^T}(A))$ or $\alpha = \textrm{rank}(\Delta_{Z_0^T\!,\, Z_0}(A))$. In the case of the Hankel operator, Equation~\eqref{eq:toep1} becomes
\begin{equation}\label{eq:hankel}
   A^{-1} = \sum_{i=1}^{\alpha} G(X_i)U(Y_i),
\end{equation}
where $G(\cdot)$ is the block Hankel matrix defined as
\begin{equation}\label{def:G}
    G(W) = 
    \begin{bmatrix}
    w_{1} & \ldots & w_{n-2} & w_{n-1} & w_{n} \\
    w_{2}& \ldots & w_{n-1} & w_{n} & 0 \\
    w_{3} & \ldots & w_{n} & 0 & 0 \\
    \vdots & \iddots & \vdots & \vdots & \vdots \\
    w_{n} & \ldots & 0 & 0 & 0
    \end{bmatrix},
\end{equation}
and Equation~\eqref{eq:lu} instead becomes the recovery formula 
\begin{equation}\label{eq:recovery2}
    [G(X_i)U(Y_i)]_{j, k} = \sum_{1 \leq \ell \leq \min\{k,\,m-j+1\}} [X_i Y_i^T]_{j+\ell-1,\, k-\ell+1}.
\end{equation}

\subsection{Upper Bound for 
Inversion of Block Toeplitz/Hankel-Like Matrices}
We begin by recalling the definition of the dual exponent:
\begin{definition}[\cite{GU18}]\label{def:dualexp}
The \textit{dual exponent} of matrix multiplication, denoted by $\beta$, is defined as the quantity $\beta = \sup \{k \mid \omega(k) = 2 \}$.
\end{definition}

In other words, $\beta$ is defined as the asymptotically maximum number $b \leq 1$ such that multiplying an $n \times n^b$ matrix by an $n^b \times n$ matrix can be done in $n^{2 + o(1)}$ time. 
In the particular case when $A$ is a blocked Toeplitz/Hankel-like matrix, 
we obtain the following:

\begin{theorem}\label{blockTH}
Given a blocked Toeplitz/Hankel-like matrix in $F^{n \times n}$ with blocking size $s = n/m$ 
and displacement rank $\alpha = n^{o(1/(\omega-1))}$,
its explicit inverse can be obtained in $\Ohat \big(s^{\omega}m + n^{\omega_s} + n^2\big) = \Ohat(n^{\omega_s})$ time. For $s < n^{\beta}$ where $\beta$ is the dual exponent, this running time becomes $O\big(n^{2 + o(1)}\big)$.
\end{theorem}

\begin{proof}
Computing the implicit inverse given by Theorem~\ref{thm:ba80} requires $\Ohat (s^{\omega}m)$ operations. Multiplying the rectangular generators requires $\Ohat \big(n^{\omega_s}\big)$ operations, and finally recovering the inverse from their product requires $O(n^2)$ operations.  
\end{proof}

\begin{remark}
The current best lower bound for $\beta$ was obtained in~\cite{GU18}, and is $\beta \geq 0.31389$. Thus our algorithm achieves exactly $\Ohat(n^{2})$ for inverting Toeplitz/Hankel-like matrices with blocking size $s$ smaller than $n^{0.31389}$.
\end{remark}

Note that $n^{\omega_s}$ as given by~\cite{GU18} is strictly smaller than $s^{\omega}m^2$ for values of $s< n$ (see Figure~\ref{fig:plot} and the proof of Theorem~\ref{thm:inv}). Thus, this improves on the best upper bound for the explicit inversion of block Toeplitz/Hankel matrices, which was $\Ohat(s^{\omega}m^2)$. This running time can be obtained by applying FFT $m$ times to the low displacement rank representation, essentially treating each column of the matrix as a separate vector. This is also the running time that Eberly et~al.\ obtain for multiplying $H^{-1}M$ with an arbitrary matrix $M$. However, their procedure noted no difference between performing the product $H^{-1}M$ or the product $L(X_i)U(Y_i)$, which occurs between matrices that are both very structured (triangular and Toeplitz/Hankel).

\begin{algorithm}\caption{Inverting a block Toeplitz/Hankel-like matrix}\label{alg:blockth}
\begin{algorithmic}[1]
\Procedure{\textsc{BlockStructInv}}{$A,s,m$}
\State For the Toeplitz operator, $P=Z_0, Q= Z_0^T$.
\State For the Hankel operator, $P=Z_0^T, Q=Z_0$. \Comment{Theorem~\ref{thm:expinv}}
    \State $\Delta_{P, Q} \leftarrow$ displacement operators associated to $A$. \Comment{Definition \ref{def:operator}}
    \State Let $X_i, Y_i$ be 
    rectangular generators such that $\Delta_{P, Q}(A^{-1}) = \sum_{i=1}^{\alpha} X_iY_i^T$.
    \State Compute $X_i, Y_i \in F^{n \times s}$. \Comment{\cite{BJS07} algorithm (Thm.\;\ref{thm:bjs})}
    \State Perform the $\alpha$ rectangular products $X_iY_i^T$. \Comment{Fast algorithm by \cite{GU18}}
    \State 
    In block Toeplitz-like case,
        \State \quad \Return  $A^{-1} = \sum_{i=1}^{\alpha} L(X_i)U(Y_i)$. \Comment{Using Equation \eqref{eq:lu}}
    \State 
    In block Hankel-like case,
    \State \quad \Return  $A^{-1} = \sum_{i=1}^{\alpha} G(X_i)U(Y_i)$. \Comment{Using Equation \eqref{eq:hankel}}
\EndProcedure
\end{algorithmic}
\end{algorithm}

\subsection{Other Displacement Operators}
The idea of recovering the explicit inverse directly from the rectangular product $XY^T$, where $X, Y \in F^{n \times \alpha s}$ are rectangular generators such that $\Delta_{P, Q}(A^{-1}) = XY^T$, extends to other kinds of matrices beyond Toeplitz/Hankel-like that also have low rank generators. This is also one of the improvements of our method over that of~\cite{EGGSV07}, since the algorithm that they use to invert the block Hankel matrix (namely the off-diagonal formula of Beckermann and Labahn) works strictly only for Hankel matrices, and does not generalize to other types of structured matrices. While most papers on
subquadratic algorithms for structured linear system solvers have focused on the Toeplitz/Hankel case, a variety of operators exist for other types of structured matrices~\cite{P01}. Well-known cases of such matrices are Vandermonde and Cauchy. We write down what the generators $P$ and $Q$ are in each case to demonstrate the claim. 


\noindent
\textbf{Vandermonde matrices.} 
In this case, besides the circulant matrix $Z_0$,
we use the following notation for diagonal and Vandermonde matrices:
\begin{equation}\label{eq:D}
D(U) = 
\begin{bmatrix}
u_1 & & &  \\
& u_2 & & \\
& & \ddots & \\
& & & u_{n}
\end{bmatrix},
\qquad
V(U) = 
\begin{bmatrix}
1 & u_{1} & u_{1}^2 & \ldots & u_{1}^{n-1} \\[0.1cm]
1 & u_{2} & u_{2}^2 & \ldots & u_{2}^{n-1} \\[0.1cm]
1 & u_{3} & u_{3}^2 & \ldots & u_{3}^{n-1} \\
\vdots & \vdots & \vdots & \ddots & \vdots \\
1 & u_{n} & u_{n}^2 & \ldots & u_{n}^{n-1} \\
\end{bmatrix}.
\end{equation}

The displacement operator for a Vandermonde matrix $V(U)$ is defined as
\[
\Delta_{D(U),\, Z_0^T}(A) = A - D(U)\, A\, Z_0^T,
\]
which yields a rank-$1$ matrix precisely when $A=V(U)$.

The analogous version of 
Theorem~\ref{3.4} for Vandermonde-like matrices with displacement rank $\alpha$ reads as follows:
\begin{theorem}[{\cite[\S 4.4]{P01}}]\label{thm:vandermonde}
Given scalars $u_1,\dots,u_n$ and generators $X, Y \in F^{n \times \alpha}$, or equivalently given generators $X_i, Y_i \in F^{n \times 1}$, where $1 \leq i \leq \alpha$, the equation 
\[
\Delta_{D(U),\, Z_0^T}(A) = A - D(U)\,A\,Z_0^T = \sum_{i=1}^{\alpha} X_i Y_i^T = XY^T
\]
has the unique solution
\begin{equation}\label{vm}
A = \sum_{i=1}^{\alpha} D(X_i) \, V(U)\, L(Y_i)^T.
\end{equation}
\end{theorem}

The analogous version of Theorem 
\ref{thm:ba80} is in turn:
\begin{theorem}[\cite{P90, BJS07}]\label{thm:rectvan}
Given a blocked input matrix $A$ with blocking size $s = n/m$, 
the rectangular generators $X_i, Y_i$ of $\Delta_{D(U),\, Z_0^T}(A^{-1})$ can be found in time $\Ohat(\alpha^{\omega-1} s^{\omega}m \log^2 m)$.
\end{theorem}

A block version of Theorem \ref{thm:vandermonde} is obtained by
letting the $u_i$ be $s\times s$ matrices.
In this case, $X,Y\in F^{n\times \alpha s}$. If $s>1$, then the recovery formula \eqref{vm} does not hold in general,
due to the non-commutativity of the parameter matrices $u_i$ with the blocks of~$X$.
However, the following decompressing relation for block Vandermonde-like matrices does hold.

\begin{lemma}
Given matrices $u_1,\dots,u_{m} \in F^{s \times s}$ and generators $X, Y \in F^{n \times \alpha s}$, where $m=n/s$, or equivalently given generators $X_i, Y_i \in F^{n \times s}$, where $1 \leq i \leq \alpha$, the equation 
\[
\Delta_{D(U),\, Z_0^T}(A) = A - D(U)\,A\,Z_0^T = \sum_{i=1}^{\alpha} X_i Y_i^T = XY^T
\]
has the unique solution
\begin{equation}\label{bvm}
\text{$A_{i, j} = \sum_{k=1}^j u_{i}^{j-k}\, [XY^T]_{i, k}$.}
\end{equation}
\end{lemma}

\begin{proof}
The equality $A-D(U)AZ_0^T=XY^T$ implies that $A_{i,1}=[XY^T]_{i,1}$ for all $i$, and
\begin{equation}\label{eqvandermonde}
A_{i,j} = [XY^T]_{i,j} + u_{i} A_{i,j-1} \ \text{for $j>1$ and every $i$.}
\end{equation}
This proves \eqref{bvm} and provides an efficient recursion for computing the entries of~$A$. The resulting matrix coincides with \eqref{vm} if $u_1,\dots,u_{m}$ commute with the components 
of~$X$. 
\end{proof}

\begin{theorem}\label{thm:blockvandermonde}
For a block Vandermonde-like $n\times n$ matrix $A$ with displacement rank $\alpha$, block size $s=n/m$ and parameter matrices $u_1,\dots,u_{m}$, the running time required to compute $A^{-1}$ explicitly is $\Ohat(\alpha^{\omega-1} n^{\omega_s} + \tau m^2)$,
where $\tau$ is the maximum cost of multiplying one of the matrices $u_i$ by an arbitrary $s\times s$ matrix.
\end{theorem}

\begin{proof}
By Theorem \ref{thm:rectvan}, we can compute the rectangular generators of $A^{-1}$ and then obtain $\Delta_{D(U),\, Z_0^T}(A^{-1})$ with $\alpha$ rectangular multiplications. 
Recovering $A^{-1}$ from $\Delta_{D(U),\, Z_0^T}(A^{-1})$ by means of \eqref{eqvandermonde} requires $m^2$ sums of $s \times s$ matrices and $m^2$ products by the matrices $u_1,\dots,u_{m}$. Thus the recovery step amounts to $\Ohat(\tau m^2)$.
\end{proof}

Note that if the matrices $u_1, \ldots, u_{m}$ are scalar multiples of the identity, or are sparse, or more generally whenever $\tau < \alpha^{\omega-1}n^{\omega_s}$, the running time required for inverting such a matrix explicitly is $\Ohat(\alpha^{\omega-1} n^{\omega_s})$. For any $\tau < s^w$ and $\alpha = o(n^{1/(\omega-1)})$, the above yields a running time improvement over \cite{EGGSV07}. The final algorithm is analogous to the one presented in Algorithm~\ref{alg:blockth}.

\subsection{General Statement}\label{sec:generalstatement}
For any block structured matrix with structure matrices $P, Q$, we can attempt to find a recursive way to reconstruct $A^{-1}$ from $\Delta_{P, Q}(A^{-1})$ in time $\Tilde{O}(\alpha n^2)$ by examining the structure of the unique solution to the corresponding functional equation. If so, then we can recover the explicit inverse in time $O(\alpha(n^2 + n^{\omega_s}))$ from its low displacement rank generators by employing fast rectangular matrix multiplication. More concretely, given an invertible blocked matrix $A \in F^{n \times n}$ with blocking size $s$ associated with a displacement operator $\Delta_{P, Q}$ such that $\Delta_{P, Q}(A^{-1}) = A^{-1} - PA^{-1}Q = XY^T$ has rank $\alpha s$ with $\alpha = o(n^{1/(\omega-1)})$, suppose that the following conditions hold:
\begin{enumerate}
    \item[\rm{1.}] There exists a fast algorithm
    for obtaining the rectangular generators $X, Y \in F^{n \times \alpha s}$ of $A^{-1}$ from~$A$.
    \item[\rm{2.}] The matrix $A^{-1}$ can be quickly recovered from $\Delta_{P,Q}(A^{-1})$. 
\end{enumerate}
Then $A^{-1}$ can be computed explicitly in the time required for the
above two operations plus an additional $\Ohat(\alpha
n^{\omega_s})$ time, where $n^{\omega_s}$ is the running time for
multiplication of an $n\times s$ matrix by an $s\times n$ one. For the
types of matrices discussed in this paper, i.e., block
Toeplitz, Hankel, and Vandermonde, the operation in
Condition~1 can be performed in $\Ohat(\alpha^{\omega-1} s^{\omega}m)$ time and the operation in
Condition~2 can be performed in $\Ohat(\alpha n^2)$ time.

In the case of block Toeplitz/Hankel-like matrices, Condition~1 is ensured by the~\cite{BA80} algorithm (with running time $\Ohat(\alpha^{\omega-1} s^{\omega} m \log^2 m)$; see Theorem~\ref{thm:bjs}), and Condition~2 is given by our construction in Theorem~\ref{thm:expinv}. In the case of block Vandermonde-like matrices, Condition~1 is given by the algorithm given in~\cite{P90} (with also running time $\Ohat(\alpha^{\omega-1} s^{\omega} m \log^2 m)$; see Theorem~\ref{thm:rectvan}), and Condition~2 follows from the recursion~\eqref{eqvandermonde}, assuming that the parameter matrices $u_1,\dots,u_{m}$ of the displacement operator are simple enough, e.g., sparse. 
Our heuristic is potentially generalizable to other block structured matrices that have an associated displacement operator, such as Cauchy, Toeplitz$+$Hankel, B\'ezout, Sylvester, Frobenius, or Loewner. We remark that for a block Cauchy matrix $C(U,V) = ((u_i - v_j)^{-1})^{m}_{i, j = 1}$, which has displacement operator $\Delta_{D(U),\, D(V)} (A) = D(U)\, A - A\, D(V) = XY^T$, the block version of the scalar recovery equation $A = \sum_{i=1}^{\alpha} D(X_i)\, C(U, V)\, D(Y_i)$ is $A_{i, j} = (u_i-v_j)^{-1}\,[XY^T]_{i, j}$. 
However, this block recovery formula requires to impose strong commutativity restrictions on the displacement parameter matrices $u_i, v_j$, such as assuming that they are scalar multiples of the identity. An example of the use of Cauchy-like systems can be found in \cite{HLS17}. We remark that 
one can transform Cauchy-like structure into Vandermonde-like and/or Toeplitz-Hankel-like structure by means of multiplying an input matrix by Vandermonde matrices \cite{P90, P17}.


\section{Rank and Nullspace Computation}\label{sec:rank}
\label{RankComputation}
As noted in Section~5 of~\cite{EGGSV07}, the algorithm we improved in Section~\ref{sec:inversion} for fast sparse matrix inversion can also be used to compute both the rank and the nullspace of a matrix over a finite field $F$. Our same running time improvement also holds.

Eberly et al.\ compute the rank and nullspace with probabilistic algorithms in two steps: first, they apply the algorithm by Kaltofen and Saunders~\cite{KS91} to compute the rank of $A$ with high probability. This algorithm first preconditions the matrix $A$ with random upper and lower triangular Toeplitz matrices $U, L \in F^{n \times n}$ and a random diagonal matrix $D \in F^{n \times n}$ and set $\Tilde{A} = UALD$, which allows them to subsequently prove that all the leading $i \times i$ minors of $\Tilde{A}$ for $1 \leq i \leq r$ are non-singular, where $r$ is the rank of~$A$. The final (deterministic) running time is as follows:
\begin{theorem}[\cite{KS91}]\label{thm:ks91}
For any matrix $A \in F^{n \times n}$, computing the rank of $A$ with high probability can be done in $\Ohat(n^2 + n \phi(n))$ operations in $F$. 
\end{theorem}
A~black-box application is a matrix-vector multiplication, which costs $\Ohat(n)$ operations for sparse matrices and also for structured matrices. The algorithm requires that the finite field $F$ has sufficiently many elements, yet this can be arranged by passing to an algebraic extension. However, the \cite{KS91} algorithm does not certify the output rank, and there is no known method to do so in the running time given in Theorem~\ref{thm:ks91} \cite{EGGSV07}.

The second step is to certify the rank obtained by the algorithm in~\cite{KS91}, which otherwise is not guaranteed to be correct. An algorithm for rank certification and nullspace computation is presented in Section~5 of Eberly et~al. The algorithm is as follows. We first partition the preconditioned matrix (which we rename as $A$) into four blocks determined by the leading (non-singular) $r \times r$ minor $A_0$:
\begin{equation}\label{eq:blocks}
    A = 
    \begin{bmatrix}
    A_0 & A_1 \\
    A_2 & A_3
    \end{bmatrix}.
\end{equation}
Next, we invert $A_0$, and here is where we apply our algorithm for 
black-box matrix inversion (Theorem~\ref{thm:inv}), which takes $O\big(n^{2.2131}\big)$ field operations, instead of the Eberly et~al.\ construction and running time. If $A_0$ is invertible, then the actual rank of $A$ is at least the estimated $r$ given by the~\cite{KS91} algorithm. 

Finally, we compute the Schur complement of the principal minor, namely $A_2A_0^{-1}A_1-A_3$, and check if it is 0. If so, we output the rank $r$ and the nullspace of $A$, which is given by
\begin{equation}\label{eq:nullspace}
    \begin{bmatrix}
    A_0^{-1}A_1 \\[0.1cm] -I
    \end{bmatrix}.
\end{equation}
To compute the Schur complement, we can no longer use the fact that we can multiply $A_0^{-1}$ efficiently with an arbitrary matrix as in Eberly et~al., since in our construction we made the inverse explicit. However, since $A$ is an efficient black box (due to the original sparsity and the structure of the preconditoning matrices $U, L, D$), we can still treat $A_1 \in F^{r \times (n-r)}$ and $A_2 \in F^{(n-r) \times r}$ as efficient black boxes.
\begin{lemma}\label{lemma:schur}
For a matrix $A \in F^{n \times n}$ with a non-singular leading minor~$A_0$, the Schur complement $A_2 A_0^{-1}A_1  - A_3$ can be computed in time 
\[
\Ohat(mn\,\phi(n) + n^{\omega_s} + mn^2 + n\,\phi(n)).
\]
If $\phi(n) = \Ohat(n)$, then an expected number of $O\big(n^{2.2131}\big)$ operations is required.
\end{lemma}
\begin{proof}
Let $r \in \Z$ be such that $A_0 \in F^{r \times r}$, and thus $A_1 \in F^{r \times (n-r)}$ and $A_2 \in F^{(n-r) \times r}$. By Theorem~\ref{thm:inv}, the time to invert $A_0$ explicitly is $\Tilde{O}(mn\,\phi(n) + n^{\omega_s} + mn^2)$. To compute $A_0^{-1}A_1$, we will instead perform the product $A_1^T(A_0^{-1})^T$ and then transpose. First consider the case where $r \leq n-r$, and divide $A_1^T \in F^{(n-r) \times r}$ into square blocks of size $r \times r$. Applying each block to $(A_0^{-1})^T$ requires at most $r\,\phi(n)$ operations, and there are $\left \lfloor{n/r}\right \rfloor$ such blocks. Therefore, computing $A_0^{-1}A_1$ requires $n$ black-box applications of $A_1$, or at most $n\,\phi(n)$ operations. In the second case, we have $r \geq n-r$. We add $2r-n$ rows of 0s to $A_1^T$ to turn it into a square matrix and then perform the product $A_1^T (A_0^{-1})^T$ with $r$ black-box applications. Overall, we require at most $n\,\phi(n)$ operations to compute $A_0^{-1}A_1$. The same construction carries over when performing the product $A_2\cdot(A_0^{-1}A_1)$, which we can obtain with at most $n$ black-box applications of $A_2$.
\end{proof}

By assembling the~\cite{KS91} algorithm for probabilistically computing the rank with the Eberly et~al.\ nullspace computation and rank certification, along with our speed-up for sparse matrix inversion, we obtain the following final running time:
\begin{theorem}\label{thm:rank}
\label{rank}
Let $A \in F^{n \times n}$ be a non-singular matrix, where $F$ is a finite field. The procedure \textsc{MatrixRankAndNullspace}$(A)$
(see Algorithm~{\rm\ref{alg:rank}}) returns the rank $r$ of $A$, and a matrix $N$ whose columns form a basis of the nullspace of $A$, in expected time
\[
\Ohat(mn\,\phi(n)+s^{\omega}m + n^{\omega_s} + mn^2),
\]
where $sm=n$. If~$\phi(n) = \Ohat(n)$, e.g., if $A$ is sparse, then the inverse $A^{-1}$ can be computed in expected time
\[
\Ohat \big(n^{\omega(k)}\big),
\]
where $k = \log_n s$ is the only value that satisfies $\omega(k) = 3-k$. With the current values of rectangular matrix multiplication, this corresponds to $O\big(n^{2.2131}\big)$ arithmetic operations.
\end{theorem}

\begin{proof}
Given a sparse matrix $A \in F^{n \times n}$, using the algorithm described in this section and Lemma~2, we can compute a basis $\{v_i\}$ of the nullspace of the preconditioned matrix $\Tilde{A} = UALD$ in time $O(n^{2.2131})$. Then $\{LDv_i\}$ is a basis of the nullspace of $A$.
Since $D$ is a diagonal matrix and $L$ is a lower triangular Toeplitz matrix, the additional multiplications only cost $O(n^2)$ and thus do not add an overhead to the running time.
\end{proof}

\begin{algorithm}\caption{Computing the rank and nullspace of an arbitrary matrix over a finite field}\label{alg:rank}
\begin{algorithmic}[1]
\Procedure{\textsc{MatrixRankAndNullspace}}{$A$} \Comment{Theorem~\ref{thm:rank}}
    \State $r \leftarrow$ rank of $A$ w.h.p. \Comment{\cite{KS91} algorithm (Thm.\;\ref{thm:ks91})}
    \State Partition $A$ into the 4 blocks $A_0, A_1, A_2, A_3$. \Comment{Equation \ref{eq:blocks}}
    \State Compute $A_0^{-1}$ (if singular, re-try the \cite{KS91} algorithm). \Comment{Alg. \ref{alg:main1} or Alg. \ref{alg:main2}}
    \State Compute the Schur complement $A_2A_0^{-1}A_1-A_3$. \Comment{Lemma \ref{lemma:schur}}
    \If{$A_2A_0^{-1}A_1-A_3 = 0$}
        \State $N \leftarrow \begin{bmatrix}
                A_0^{-1}A_1 | -I
                \end{bmatrix}^T$
                \Comment{Nullspace of $A$}
        \State \Return $r,N$.
    \Else
        \State Restart from Line 2.
    \EndIf
    \EndProcedure
\end{algorithmic}
\end{algorithm}


More generally, the rank and nullspace algorithm presented in this section yield the following observation:
\begin{remark}
For any matrix inversion algorithm \textsc{MatrixInv} with running time $\textsc{T}_{\textsc{MatrixInv}}$, the algorithm \textsc{MatrixRankAndNullspace} requires an expected running time of
\[
\textsc{T}_{\textsc{MatrixRankAndNullspace}} = \textsc{T}_{\textsc{MatrixInv}} + n\,\phi(n).
\]
In particular, for sparse matrices the running time of the two algorithms is the same.
\end{remark}

%

\section{Applications}
The algorithm described in Section~\ref{RankComputation} for computation of the rank and nullspace of a sparse matrix over a finite field has multiple applications. Some relevant to theoretical computer science include efficient decoding of algebraic-geometric codes \cite{JM89, OS99}, low density parity check codes \cite{BCH10}, discrete logarithm computations in cryptography \cite{JP16}, and (multivariate) polynomial interpolation \cite{O06}.

\subsection{Topological Data Analysis}
\label{TDA}

Our first detailed example deals with the calculation of persistent homology in topological data analysis.
Persistent homology is a widely used technique, based on algebraic topology, to determine shape features of point clouds \cite{ELZ00,ZC05,EH08}.
To a point cloud $X$ (i.e., a finite set of points in Euclidean space) one associates a filtered simplicial complex
$V(X)=\{V_{\varepsilon}(X)\}_{\varepsilon>0}$ by one of several possible methods \cite{OPTGH17}; for instance, the \emph{Vietoris--Rips complex}
contains, for each value of~$\varepsilon$, a $k$-simplex for each set of $k+1$ points in $X$ with diameter less than or equal to~$\varepsilon$. 
The \emph{persistence diagram} of $X$ has a point $(b,d)$ with $d>b$ for each homology generator in any dimension of $V(X)$ arising at a parameter value $\varepsilon=b$ (birth) and vanishing at~$\varepsilon=d$ (death). The \emph{persistence} or \emph{lifetime} of such a homology class is then defined to be $d-b$. A~non-zero homology class in dimension $k$ is represented by a $k$-cycle that is not a boundary of any chain of $(k+1)$-simplices. 

Computing a persistence diagram for a point cloud $X$ requires finding ranks of matrices of boundary operators on~$V(X)$; see \cite[\S\,5.3]{OPTGH17}. For convenience, we assume that coefficients in the field $F=\mathbb{Z}/2\mathbb{Z}$ are used. 
The maximum number of linearly independent homology classes in dimension $k$ of a simplicial complex is called the \emph{$k$-th Betti number} of that complex. The running time of algorithms based on Gaussian Elimination for the calculation of Betti numbers and persistent homology is $O(n^3)$ where $n$ is the total number of simplices in the given complex. Using less straight\-forward methods, the complexity was reduced to $O(n^{\omega})$ in~\cite{MMS11}.

An output-sensitive algorithm for the computation of persistence diagrams was described in~\cite{CK13} with the following running time.

\begin{theorem}[\cite{CK13}]
Given a filtered simplicial complex with $n$ simplices, let $C_{\Gamma}$ denote the number of homology generators with persistence at least $\Gamma$ for any threshold $\Gamma > 0$. Then, a persistence diagram over $\mathbb{Z}/2\mathbb{Z}$ can be computed deterministically in $O(C_{\Gamma\,}n^{\omega}\log n)$ time or probabilistically in expected $O(C_{\Gamma}\, n^{3-1/(\omega - 1)})$ time. 
\end{theorem}

Imposing that $C_{\Gamma}$ be at most of the order of $\log n$ is a reasonable assumption in practice, because homology classes with small persistence are treated as noise in most applications of topological data analysis. Since $3-1/(\omega-1)=2.2716$ for the current value of~$\omega$, the following theorem improves the running time obtained in~\cite{CK13}. This constitutes an interesting application of our results in which the sparsity of the matrix is inherent and does not need to be imposed. 

\begin{theorem}\label{thm:homology}
For a filtered simplicial complex with a total number of $n$ simplices, a persistence diagram over $\mathbb{Z}/2\mathbb{Z}$ can be computed in expected time $O\big(n^{2.2131}\big)$ if the persistence of homology generators is bounded below so that their number is at most logarithmic in~$n$.
\end{theorem}

\begin{proof}
The algorithm of \cite{CK13} requires computing ranks of a certain collection of sparse submatrices of an $n\times n$ matrix, which can be done with $O\big(n^{2.2131}\big)$ field operations by Theorem~\ref{rank}.
\end{proof}

We also provide the following complexity for ordinary (not persistent) homology. Although this result also applies to simplicial complexes equipped with a filter function, its randomized nature obstructs the precise determination of persistence of cycles.

\begin{theorem}\label{thm:onecomplex}
If a simplicial complex $V$ has a total number of $n$ simplices, then the Betti numbers of $V$ with coefficients in a finite field
together with a basis of cycles in each dimension can be computed in expected time $O\big(n^{2.2131}\big)$.
\end{theorem}

\begin{proof} 
The matrices of boundary operators on $V$ can be assembled into a single $n\times n$ matrix $A$ 
such that $A_{i,j}=\pm 1$ if and only if the $i$-th simplex occurs in the boundary of the $j$-th simplex. Hence this matrix $A$ is upper triangular and has $k+1$ non-zero entries in each column corresponding to a $k$-simplex. Since dimension depends logarithmically on the number of simplices (because a $k$-simplex has $2^{k+1}-1$ faces), the total number of non-zero entries in $A$ is $O(n\log n)$. Consequently, $A$ is sparse.

It then follows from Theorem~\ref{rank} that the rank and nullspace of $A$ can be computed with an expected number of $O\big(n^{2.2131}\big)$ arithmetic operations. The rank of the boundary operator on each dimension can be obtained similarly by restricting the calculation to the corresponding submatrix. Knowledge of the nullspace of $A$ yields a set of linearly independent generating cycles in each dimension.
Since the number of dimensions is $O(\log n)$, our claim follows.
\end{proof}

\subsection{Units in Group Rings}
For a finite group $G$ of order $n$ and a ring $R$ with unity, the group ring $R[G]$ is isomorphic to a certain subring of the ring of $n\times n$ matrices over~$R$, as proved in~\cite{H06}. If we assume that the elements of $G$ are ordered as $g_1,\dots,g_n$, then we may consider the matrix $M(G)=\left(g_i^{-1}g_j\right)$ for $i=1,\dots,n$ and $j=1,\dots,n$. The elements in the group ring $R[G]$ are formal sums of elements $g_i \in G$ with coefficients $\beta_{g_i} \in R$. For each element $\beta=\sum_{i=1}^n\beta_{g_i}\,g_i$ in $R[G]$ we are concerned with the problem of determining whether a given element $\beta$ is a unit or not. 

Following the method outlined in \cite{GJS20}, to each element $\beta \in R[G]$ we may assign the matrix $M_\beta=\big(\beta_{g_i^{-1}g_j}\big)$.
This is the matrix of right-multiplication by $\beta$ written in the $R$-basis $g_1,\dots,g_n$. Hence $\beta\mapsto M_\beta$ sets up an injective ring homomorphism \cite[Theorem~1]{H06}. This yields a method to test if a given element $\beta\in R[G]$ is a unit, by checking if the matrix $M_\beta$ is invertible. 
In fact, if $\beta$ happens to be a unit, then $\beta^{-1}$ is associated with the inverse matrix~$M_\beta^{-1}$.

As in \cite[\S\,4.2]{GJS20}, we consider the case when the group $G$ is metacyclic and coefficients in a field $F$ are used. We will assume, however, that the field $F$ is finite. When $G$ is metacyclic, it admits a presentation of the form
\[
\langle\sigma,\tau\mid \sigma^m=1,\,\tau^s=\sigma^t,\,\tau^{-1}\sigma\tau=\sigma^n\rangle
\]
for integers $m,t,u,s$ with $u\le m$, $t\le m$, $u^s\equiv 1$ mod~$t$, and $ut\equiv t$ mod~$m$. The order of $G$ is $n=ms$. If we list the elements of $G$ as 
\[
1,\tau,\dots,\tau^{s-1},\sigma, \sigma\tau,\dots,\sigma\tau^{s-1},\dots,\sigma^{m-1},\sigma^{m-1}\tau,\dots,\sigma^{m-1}\tau^{s-1}
\]
then the matrix $M(G)$ is block Toeplitz with blocks of size $s\times s$. Therefore each matrix $M_\beta$ for $\beta\in F[G]$ is also block Toeplitz, so its invertibility can be tested with $\Tilde{O}(s^{\omega}m)$ operations in~$F$. As observed in \cite[\S\,4.2]{GJS20}, it is equally possible to exchange the roles of $\sigma$ and $\tau$ so that we obtain a block Toeplitz matrix with blocks of size $m\times m$ instead.

\begin{theorem}\label{thm:metacyclic}
For a finite field $F$ and a finite group $G$ with a normal subgroup which is cyclic of order $m$ and cyclic quotient of order~$s$, determining if an element $\beta\in F[G]$ is invertible and, if so, computing the set $\{g\beta^{-1}\}_{g\in G}$ explicitly can be done in 
expected time $O(\omega(\min\{\log_n m, \log_n s\}))$.
\end{theorem}

\begin{proof} The $G$-orbit $\{g\beta^{-1}\}_{g\in G}$ corresponds to the rows of the inverse matrix~$M_\beta^{-1}$. Since $M_\beta$ is block Toeplitz, the result follows from Theorem~\ref{blockTH}. 
\end{proof}

Theorem~\ref{thm:metacyclic} yields approximately quadratic running time, since one of $m$ or $s$ is smaller than or equal to~$\sqrt{n}$, and $\omega(0.5)=2.0442$ according to~\cite{GU18}. For a field $F$ of characteristic zero, the running time given in \cite[Proposition~4.13]{GJS20} is $\Ohat(n^{(\omega+1)/2})$ in order to check invertibility of $M_\beta$ and obtain~$\beta^{-1}$, while the $G$-orbit of $\beta$ can be written down in $\Ohat(n^{(\omega+3)/2})=O(n^{2.6864})$.


%

\section{Open Problems}
A central open question that our work does not resolve is whether a similar running time improvement for sparse matrix inversion as the one given in this paper for finite fields can also be obtained in finite precision arithmetic. We have improved the running time in the finite field setting to $O(n^{2.2131})$, but in finite precision arithmetic there is still no known sub-matrix-multiplication algorithm for this problem, and the current best running time is still $\Ohat(n^{\omega})$.

In contrast, \cite{PV21} recently obtained an $O(n^{2.3316}) < \Ohat(n^{\omega})$ time algorithm in finite precision arithmetic for sparse linear system solving. If we attempt to apply our construction of low displacement rank methods paired up with fast rectangular matrix multiplication to the numerical analysis of \cite{PV21}, the bottleneck occurs when multiplying the two rectangular $n \times s$ matrices that represent $H^{-1}$, which each now require $\Ohat(m \log\kappa)$ words of precision (where $\kappa$ is the condition number of the matrix). This rectangular product would then cost $mn^{\omega_s}$, and since $mn^{\omega_s} \geq n^{\omega}$ (as shown in Section~\ref{sec:hinv}), our algorithm is not able to beat matrix multiplication time in the finite precision setting.

Another interesting question is whether our method can be used to improve the Eberly et~al. running time of $\Otil(n^{2.5})$ for solving sparse linear systems exactly over the rationals, and lowering it below matrix multiplication time. Their algorithm uses Dixon's rational reconstruction algorithms and $p$-adic numbers \cite{D82}. The reason for the inapplicability of our construction is that in making the block Hankel inverse matrix explicit in the construction by Eberly et~al., we are then unable to apply a vector efficiently to $H^{-1}$. This vector application is required in each of the $n$ iterations of Dixon's algorithm. Nonetheless, we believe that it should be possible to improve Eberly et~al.'s $\Tilde{O}(n^{2.5})$ running time for sparse linear system solving by also blocking the solution $x$ and then using polynomial multiplication \cite{CK91}. In other words, by obtaining $s$ $p$-adic digits of the solution $x$ per iteration and thus requiring only $m = n/s$ Dixon iterations instead of $n$ (or, equivalently, to perform Eberly et~al.'s algorithm in $\Z/p^s\Z$ rather than in $\Z/p\Z$). In particular, blocking the $p$-adic solution $x$ appears to be a useful approach given the recent developments in relaxed $p$-adic arithmetic~\cite{BHL11, BL12}, which significantly reduced the running time needed to multiply two $p$-adic expressions.

\addcontentsline{toc}{section}{Acknowledgements}
\section*{Acknowledgements}
We are thankful to Richard Peng and Markus P\"uschel for helpful suggestions and comments. We are also grateful to Romain Lebreton and Victor Pan for pointing out corrections and for their interest in our work.


\printbibliography

@inproceedings{W12,
	title = {Multiplying matrices faster than {Coppersmith}-{Wino}-grad},
	booktitle = {Proceedings of the 44th annual {ACM} {Symposium} on {Theory} of {Computing} ({STOC})},
	author = {Vassilevska Williams, Virginia},
	year = {2012},
	pages = {887--898},
}

@inproceedings{AW21,
	title = {A refined laser method and faster matrix multiplication},
	booktitle = {Proceedings of the 2021 {ACM}-{SIAM} {Symposium} on {Discrete} {Algorithms} ({SODA})},
	publisher = {SIAM},
	author = {Alman, Josh and Vassilevska Williams, Virginia},
	year = {2021},
	pages = {522--539},
}

@inproceedings{LG14,
	title = {Powers of tensors and fast matrix multiplication},
	booktitle = {Proceedings of the 39th {International} {Symposium} on {Symbolic} and {Algebraic} {Computation} ({ISSAC})},
	author = {Gall, François Le},
	year = {2014},
	pages = {296--303},
}

@article{S69,
	title = {Gaussian elimination is not optimal},
	volume = {13},
	number = {4},
	journal = {Numerische Mathematik},
	author = {Strassen, Volker},
	year = {1969},
	note = {Publisher: Springer},
	pages = {354--356},
}

@inproceedings{PV21,
	title = {Solving sparse linear systems faster than matrix multiplication},
	booktitle = {Proceedings of the 2021 {ACM}-{SIAM} {Symposium} on {Discrete} {Algorithms} ({SODA})},
	publisher = {SIAM},
	author = {Peng, Richard and Vempala, Santosh},
	year = {2021},
	pages = {504--521},
}

@article{GS72,
	title = {On the inversion of finite {Toeplitz} matrices and their continuous analogs},
	volume = {7},
	number = {12},
	journal = {Mat. Issled.},
	author = {Gohberg, Israel and Semencul, Arkadii},
	year = {1972},
	pages = {201--233},
}

@article{GK72,
	title = {A formula for the inversion of finite {Toeplitz} matrices},
	volume = {7},
	number = {2},
	journal = {Mat. Issled.},
	author = {Gohberg, Israel and Krupnik, Naum Ya},
	year = {1972},
	pages = {272--283},
}

@article{HP98,
	title = {Fast rectangular matrix multiplication and applications},
	volume = {14},
	number = {2},
	journal = {Journal of Complexity},
	author = {Huang, Xiaohan and Pan, Victor Y.},
	year = {1998},
	note = {Publisher: Elsevier},
	pages = {257--299},
}

@article{EH08,
	title = {Persistent homology - a survey},
	volume = {453},
	journal = {Contemporary Mathematics},
	author = {Edelsbrunner, Herbert and Harer, John},
	year = {2008},
	note = {Publisher: Providence, RI: American Mathematical Society},
	pages = {257--282},
}

@article{CK91,
	title = {On fast multiplication of polynomials over arbitrary algebras},
	volume = {28},
	journal = {Acta Informatica},
	author = {Cantor, David G. and Kaltofen, Erich},
	year = {1991},
	pages = {693--701},
}

@inproceedings{BJS07,
	title = {Solving {Toeplitz}- and {Van}-dermonde-like linear systems with large displacement rank},
	booktitle = {Proceedings of the 2007 {International} {Symposium} on {Symbolic} and {Algebraic} {Computation} ({ISSAC})},
	author = {Bostan, Alin and Jeannerod, Claude-Pierre and Schost, Éric},
	year = {2007},
	pages = {33--40},
}

@inproceedings{BL12,
	title = {Relaxed $p$-adic {Hensel} lifting for algebraic systems},
	booktitle = {Proceedings of the 37th {International} {Symposium} on {Symbolic} and {Algebraic} {Computation} ({ISSAC})},
	author = {Berthomieu, Jérémy and Lebreton, Romain},
	year = {2012},
	pages = {59--66},
}

@article{JM89,
	title = {A simple {Hankel} interpretation of the {Berle}-kamp-{Massey} algorithm},
	volume = {125},
	journal = {Linear Algebra and its Applications},
	author = {Jonckheere, Edmund and Ma, Chingwo},
	year = {1989},
	note = {Publisher: Elsevier},
	pages = {65--76},
}

@inproceedings{GJV03,
	title = {On the complexity of polynomial matrix computations},
	booktitle = {Proceedings of the 2003 {International} {Symposium} on {Symbolic} and {Algebraic} {Computation} ({ISSAC})},
	author = {Giorgi, Pascal and Jeannerod, Claude-Pierre and Villard, Gilles},
	year = {2003},
	pages = {135--142},
}

@inproceedings{OS99,
	title = {A displacement approach to efficient decoding of algebraic-geometric codes},
	booktitle = {Proceedings of the 31st annual {ACM} {Symposium} on {Theory} of {Computing} ({STOC})},
	author = {Olshevsky, Vadim and Shokrollahi, Amin},
	year = {1999},
	pages = {235--244},
}

@article{W86,
	title = {Solving sparse linear equations over finite fields},
	volume = {32},
	number = {1},
	journal = {IEEE Transactions on Information Theory},
	author = {Wiedemann, Douglas},
	year = {1986},
	note = {Publisher: IEEE},
	pages = {54--62},
}

@inproceedings{KZ17,
	title = {Hardness {Results} for {Structured} {Linear} {Systems}},
	booktitle = {2017 {IEEE} 58th {Annual} {Symposium} on {Foundations} of {Computer} {Science} ({FOCS})},
	publisher = {IEEE},
	author = {Kyng, Rasmus and Zhang, Peng},
	year = {2017},
	pages = {684--695},
}

@inproceedings{KWZ20,
	title = {Packing {LPs} are hard to solve accurately, assuming linear equations are hard},
	booktitle = {Proceedings of the 14th {Annual} {ACM}-{SIAM} {Symposium} on {Discrete} {Algorithms} ({SODA})},
	publisher = {SIAM},
	author = {Kyng, Rasmus and Wang, Di and Zhang, Peng},
	year = {2020},
	pages = {279--296},
}

@inproceedings{GU18,
	title = {Improved rectangular matrix multiplication using powers of the {Coppersmith}-{Winograd} tensor},
	booktitle = {Proceedings of the {Twenty}-{Ninth} {Annual} {ACM}-{SIAM} {Symposium} on {Discrete} {Algorithms} ({SODA})},
	publisher = {SIAM},
	author = {Gall, François Le and Urrutia, Florent},
	year = {2018},
	pages = {1029--1046},
}

@inproceedings{CU13,
	title = {Fast matrix multiplication using coherent configurations},
	booktitle = {Proceedings of the 24th annual {ACM}-{SIAM} {Symposium} on {Discrete} {Algorithms} ({SODA})},
	publisher = {SIAM},
	author = {Cohn, Henry and Umans, Christopher},
	year = {2013},
	pages = {1074--1087},
}

@inproceedings{EGGSV07,
	title = {Faster inversion and other black box matrix computations using efficient block projections},
	booktitle = {Proceedings of the 2007 {International} {Symposium} on {Symbolic} and {Algebraic} {Computation} ({ISSAC})},
	author = {Eberly, Wayne and Giesbrecht, Mark and Giorgi, Pascal and Storjohann, Arne and Villard, Gilles},
	year = {2007},
	pages = {143--150},
}

@inproceedings{EGGSV06,
	title = {Solving sparse rational linear systems},
	booktitle = {Proceedings of the 2006 {International} {Symposium} on {Symbolic} and {Algebraic} {Computation} ({ISSAC})},
	author = {Eberly, Wayne and Giesbrecht, Mark and Giorgi, Pascal and Storjohann, Arne and Villard, Gilles},
	year = {2006},
	pages = {63--70},
}

@inproceedings{MMS11,
	title = {Zigzag persistent homology in matrix multiplication time},
	booktitle = {Proceedings of the 27th {Annual} {Symposium} on {Computational} {Geometry}},
	author = {Milosavljević, Nikola and Morozov, Dmitriy and Skraba, Primoz},
	year = {2011},
	pages = {216--225},
}

@article{BHL11,
	title = {Relaxed algorithms for $p$-adic numbers},
	volume = {23},
	number = {3},
	journal = {Journal de théorie des nombres de Bordeaux},
	author = {Berthomieu, Jérémy and Van Der Hoeven, Joris and Lecerf, Grégoire},
	year = {2011},
	pages = {541--577},
}

@inproceedings{ELZ00,
	title = {Topological persistence and simplification},
	booktitle = {Proceedings 41st {Annual} {Symposium} on {Foundations} of {Computer} {Science} ({FOCS})},
	publisher = {IEEE},
	author = {Edelsbrunner, Herbert and Letscher, David and Zomorodian, Afra},
	year = {2000},
	pages = {454--463},
}

@article{CLS21,
	title = {Solving linear programs in the current matrix multiplication time},
	volume = {68},
	number = {1},
	journal = {Journal of the ACM (JACM)},
	author = {Cohen, Michael B. and Lee, Yin Tat and Song, Zhao},
	year = {2021},
	note = {Publisher: ACM New York, NY, USA},
	pages = {1--39},
}

@article{ST14,
	title = {Nearly linear time algorithms for preconditioning and solving symmetric, diagonally dominant linear systems},
	volume = {35},
	number = {3},
	journal = {SIAM Journal on Matrix Analysis and Applications},
	author = {Spielman, Daniel A. and Teng, Shang-Hua},
	year = {2014},
	note = {Publisher: SIAM},
	pages = {835--885},
}

@incollection{JP16,
	title = {Nearly sparse linear algebra and application to discrete logarithms computations},
	booktitle = {Contemporary {Developments} in {Finite} {Fields} and {Applications}},
	publisher = {World Scientific},
	author = {Joux, Antoine and Pierrot, Cécile},
	year = {2016},
	pages = {119--144},
}

@article{GJS20,
	title = {Subquadratic-{Time} {Algorithms} for {Normal} {Bases}},
	journal = {arXiv preprint arXiv:2005.03497},
	author = {Giesbrecht, Mark and Jamshidpey, Armin and Schost, Éric},
	year = {2020},
}

@article{OPTGH17,
	title = {A roadmap for the computation of persistent homology},
	volume = {6},
	journal = {EPJ Data Science},
	author = {Otter, Nina and Porter, Mason A. and Tillmann, Ulrike and Grindrod, Peter and Harrington, Heather A.},
	year = {2017},
	note = {Publisher: Springer},
	pages = {1--38},
}

@inproceedings{MMS18,
	title = {Stability of the {Lanczos} method for matrix function approximation},
	booktitle = {Proceedings of the {Twenty}-{Ninth} {Annual} {ACM}-{SIAM} {Symposium} on {Discrete} {Algorithms}},
	publisher = {SIAM},
	author = {Musco, Cameron and Musco, Christopher and Sidford, Aaron},
	year = {2018},
	pages = {1605--1624},
}

@article{GH74,
	title = {Inversion of finite {Toeplitz} matrices made of elements of a non-commutative algebra},
	volume = {XIX},
	number = {5},
	journal = {Rev. Roumaine Math. Pures Appl.},
	author = {Gohberg, Israel and Heinig, Georg},
	year = {1974},
	pages = {623--663},
}

@article{C94,
	title = {Solving homogeneous linear equations over {GF}(2) via block {Wiedemann} algorithm},
	volume = {62},
	number = {205},
	journal = {Mathematics of Computation},
	author = {Coppersmith, Don},
	year = {1994},
	pages = {333--350},
}

@article{D82,
	title = {Exact solution of linear equations using $p$-adic expansions},
	volume = {40},
	number = {1},
	journal = {Numerische Mathematik},
	author = {Dixon, John D.},
	year = {1982},
	note = {Publisher: Springer},
	pages = {137--141},
}

@inproceedings{S86,
	title = {The asymptotic spectrum of tensors and the exponent of matrix multiplication},
	booktitle = {27th {Annual} {Symposium} on {Foundations} of {Computer} {Science} ({FOCS} 1986)},
	publisher = {IEEE},
	author = {Strassen, Volker},
	year = {1986},
	pages = {49--54},
}

@book{P01,
	title = {Structured matrices and polynomials: unified superfast algorithms},
	shorttitle = {Structured matrices and polynomials},
	publisher = {Springer Science \& Business Media},
	author = {Pan, Victor},
	year = {2001},
}

@book{S03,
	title = {Iterative methods for sparse linear systems},
	publisher = {SIAM},
	author = {Saad, Yousef},
	year = {2003},
}

@article{ZC05,
	title = {Computing persistent homology},
	volume = {33},
	number = {2},
	journal = {Discrete \& Computational Geometry},
	author = {Zomorodian, Afra and Carlsson, Gunnar},
	year = {2005},
	note = {Publisher: Springer},
	pages = {249--274},
}

@article{S05,
	title = {The shifted number system for fast linear algebra on integer matrices},
	volume = {21},
	number = {4},
	journal = {Journal of Complexity},
	author = {Storjohann, Arne},
	year = {2005},
	note = {Publisher: Elsevier},
	pages = {609--650},
}

@article{Z02,
	title = {All pairs shortest paths using bridging sets and rectangular matrix multiplication},
	volume = {49},
	number = {3},
	journal = {Journal of the ACM (JACM)},
	author = {Zwick, Uri},
	year = {2002},
	note = {Publisher: ACM New York, NY, USA},
	pages = {289--317},
}

@article{BCH10,
	title = {Low-density parity-check codes and their rateless relatives},
	volume = {13},
	number = {1},
	journal = {IEEE Communications Surveys \& Tutorials},
	author = {Bonello, Nicholas and Chen, Sheng and Hanzo, Lajos},
	year = {2010},
	note = {Publisher: IEEE},
	pages = {3--26},
}

@article{K95,
	title = {Analysis of {Coppersmith}’s block {Wiedemann} algorithm for the parallel solution of sparse linear systems},
	volume = {64},
	number = {210},
	journal = {Mathematics of Computation},
	author = {Kaltofen, Erich},
	year = {1995},
	pages = {777--806},
}

@article{DDHK07,
	title = {Fast matrix multiplication is stable},
	volume = {106},
	number = {2},
	journal = {Numerische Mathematik},
	author = {Demmel, James and Dumitriu, Ioana and Holtz, Olga and Kleinberg, Robert},
	year = {2007},
	note = {Publisher: Springer},
	pages = {199--224},
}

@article{H06,
	title = {Group rings and rings of matrices},
	volume = {31},
	number = {3},
	journal = {Int. J. Pure Appl. Math},
	author = {Hurley, Ted},
	year = {2006},
	pages = {319--335},
}

@article{O06,
	title = {On multivariate interpolation},
	volume = {116},
	number = {2},
	journal = {Studies in Applied Mathematics},
	author = {Olver, Peter J.},
	year = {2006},
	note = {Publisher: Wiley Online Library},
	pages = {201--240},
}

@article{CK13,
	title = {An output-sensitive algorithm for persistent homology},
	volume = {46},
	number = {4},
	journal = {Computational Geometry},
	author = {Chen, Chao and Kerber, Michael},
	year = {2013},
	note = {Publisher: Elsevier},
	pages = {435--447},
}

@article{P90,
	title = {On computations with dense structured matrices},
	volume = {55},
	number = {191},
	journal = {Mathematics of Computation},
	author = {Pan, Victor},
	year = {1990},
	pages = {179--190},
}

@article{T64,
	title = {An algorithm for the inversion of finite {Toeplitz} matrices},
	volume = {12},
	number = {3},
	journal = {Journal of the Society for Industrial and Applied Mathematics},
	author = {Trench, William F.},
	year = {1964},
	note = {Publisher: SIAM},
	pages = {515--522},
}

@article{W61,
	title = {Error analysis of direct methods of matrix inversion},
	volume = {8},
	number = {3},
	journal = {Journal of the ACM (JACM)},
	author = {Wilkinson, James Hardy},
	year = {1961},
	note = {Publisher: ACM New York, NY, USA},
	pages = {281--330},
}

@article{BL94,
	title = {A uniform approach for the fast computation of matrix-type {Padé} approximants},
	volume = {15},
	number = {3},
	journal = {SIAM Journal on Matrix Analysis and Applications},
	author = {Beckermann, Bernhard and Labahn, George},
	year = {1994},
	note = {Publisher: SIAM},
	pages = {804--823},
}

@inproceedings{LO90,
	title = {Solving large sparse linear systems over finite fields},
	booktitle = {Conference on the {Theory} and {Application} of {Cryptography}},
	publisher = {Springer},
	author = {LaMacchia, Brian A. and Odlyzko, Andrew M.},
	year = {1990},
	pages = {109--133},
}

@article{LCC90,
	title = {The inverses of block {Hankel} and block {Toeplitz} matrices},
	volume = {19},
	number = {1},
	journal = {SIAM Journal on Computing},
	author = {Labahn, George and Choi, Dong Koo and Cabay, Stan},
	year = {1990},
	note = {Publisher: SIAM},
	pages = {98--123},
}

@book{HS52,
	title = {Methods of conjugate gradients for solving linear systems},
	volume = {49},
	number = {1},
	publisher = {NBS Washington, DC},
	author = {Hestenes, Magnus Rudolph and Stiefel, Eduard},
	year = {1952},
}

@article{CW82,
	title = {On the asymptotic complexity of matrix multiplication},
	volume = {11},
	number = {3},
	journal = {SIAM Journal on Computing},
	author = {Coppersmith, Don and Winograd, Shmuel},
	year = {1982},
	note = {Publisher: SIAM},
	pages = {472--492},
}

@article{P80,
	title = {New fast algorithms for matrix operations},
	volume = {9},
	number = {2},
	journal = {SIAM Journal on Computing},
	author = {Pan, Victor},
	year = {1980},
	note = {Publisher: SIAM},
	pages = {321--342},
}

@article{BA80,
	title = {Asymptotically fast solution of {Toeplitz} and related systems of linear equations},
	volume = {34},
	journal = {Linear Algebra and its Applications},
	author = {Bitmead, Robert R. and Anderson, Brian DO},
	year = {1980},
	note = {Publisher: Elsevier},
	pages = {103--116},
}

@article{LC89,
	title = {Matrix {Padé} fractions and their computation},
	volume = {18},
	number = {4},
	journal = {SIAM Journal on Computing},
	author = {Labahn, George and Cabay, Stan},
	year = {1989},
	note = {Publisher: SIAM},
	pages = {639--657},
}

@article{KKM79,
	title = {Displacement ranks of matrices and linear equations},
	volume = {68},
	number = {2},
	journal = {Journal of Mathematical Analysis and Applications},
	author = {Kailath, Thomas and Kung, Sun-Yuan and Morf, Martin},
	year = {1979},
	note = {Publisher: Elsevier},
	pages = {395--407},
}

@book{L50,
	title = {An iteration method for the solution of the eigenvalue problem of linear differential and integral operators},
	publisher = {United States Governm. Press Office Los Angeles, CA},
	author = {Lanczos, Cornelius},
	year = {1950},
}

@article{HP20,
	title = {Laplacians are {Complete} for {Linear} {System} over {Zp}},
	author = {Huang, Yufan and Peng, Richard},
	year = {2020},
}

@inproceedings{KS91,
	title = {On {Wiedemann}'s method of solving sparse linear systems},
	booktitle = {International {Symposium} on {Applied} {Algebra}, {Algebraic} {Algorithms}, and {Error}-{Correcting} {Codes}},
	publisher = {Springer},
	author = {Kaltofen, Erich and Saunders, B. David},
	year = {1991},
	pages = {29--38},
}

@inproceedings{K94,
  title={Asymptotically fast solution of Toeplitz-like singular linear systems},
  author={Kaltofen, Erich},
  booktitle={Proceedings of the international symposium on Symbolic and algebraic computation},
  pages={297--304},
  year={1994}
}

@inproceedings{HLS17,
  title={Algorithms for structured linear systems solving and their implementation},
  author={Hyun, Seung Gyu and Lebreton, Romain and Schost, {\'E}ric},
  booktitle={Proceedings of the 2017 ACM on International Symposium on Symbolic and Algebraic Computation},
  pages={205--212},
  year={2017}
}

@article{P17,
  title={Fast approximate computations with Cauchy matrices and polynomials},
  author={Pan, Victor},
  journal={Mathematics of Computation},
  volume={86},
  number={308},
  pages={2799--2826},
  year={2017}
}

\end{document}